\documentstyle[12pt]{article}
\begin{document}\thispagestyle{empty}\begin{flushright}
OUT--4102--76\\
CECM--98--120\\
hep-th/9811173\\
19 November 1998
                             \end{flushright}\vspace*{2mm}\begin{center}{
                                                           \large\bf
Determinations of rational Dedekind-zeta invariants of     \\[5pt]
hyperbolic manifolds and Feynman knots and links            \vglue 10mm
J.~M.~Borwein$^{a)}$ and D.~J.~Broadhurst$^{b)}$        }\end{center}
                                                        \vfill\noindent
                                                        {\bf Abstract}\quad
We identify 998 closed hyperbolic 3-manifolds whose volumes are rationally
related to Dedekind zeta values, with coprime integers $a$ and $b$ giving
$$\frac{a}{b}\,{\rm vol}({\cal M})=\frac{(-D)^{3/2}}{(2\pi)^{2n-4}}\,\frac
{\zeta_K(2)}{2\zeta(2)}$$ for a manifold ${\cal M}$ whose invariant trace
field $K$ has a single complex place, discriminant $D$, degree $n$, and
Dedekind zeta value $\zeta_K(2)$. The largest numerator of the 998 invariants
of Hodgson--Weeks manifolds is, astoundingly, $a=2^4\times23\times37\times691
=9,408,656$; the largest denominator is merely $b=9$. We also study the
rational invariant $a/b$ for single-complex-place cusped manifolds,
complementary to knots and links, both within and beyond the Hildebrand--Weeks
census. Within the censi, we identify 152 distinct Dedekind zetas rationally
related to volumes. Moreover, 91 census manifolds have volumes reducible to
pairs of these zeta values. Motivated by studies of Feynman diagrams, we find
a 10-component 24-crossing link in the case $n=2$ and $D=-20$. It is one of 5
alternating platonic links, the other 4 being quartic. For 8 of 10 quadratic
fields distinguished by rational relations between Dedekind zeta values and
volumes of Feynman orthoschemes, we find corresponding links. Feynman links
with $D=-39$ and $D=-84$ are missing; we expect them to be as beautiful as
the 8 drawn here. Dedekind-zeta invariants are obtained for knots from Feynman
diagrams with up to 11 loops. We identify a sextic 18-crossing positive
Feynman knot whose rational invariant, $a/b=26$, is 390 times that of the
cubic 16-crossing non-alternating knot with maximal $D_9$ symmetry. Our
results are secure, numerically, yet appear very hard to prove by analysis.
\vfill\footnoterule\noindent
$^a$) CECM, Simon Fraser University, Burnaby, B.C.\ V5A 1S6, Canada;\\
{\tt jborwein@cecm.sfu.ca;
http://www.cecm.sfu.ca/$\,\widetilde{}\,$jborwein}\\
$^b$) Physics Department, Open University, Milton Keynes MK7 6AA, UK;\\
{\tt D.Broadhurst@open.ac.uk;
http://physics.open.ac.uk/$\,\widetilde{}\,$dbroadhu}
\newpage\setcounter{page}{1}
\newcommand{\df}[2]{\mbox{$\frac{#1}{#2}$}}
\newcommand{\tabw}[5]{#1&#2&#3&#4&#5\\}
\newcommand{\tabb}[1]{#1&&&&\\\quad{}}
\newcommand{\bet}[2]{\begin{center}{\bf Table #1:}\quad#2\end{center}%
 $$\begin{array}{|l|r|rl|r|}\hline%
 \mbox{field}&-D&a&b&\mbox{manifold}\\[2pt]\hline}
\newcommand{\ent}{\hline\end{array}$$}

\section{Introduction}

This paper began with a study of hyperbolic links whose complementary
volumes result from evaluations of Feynman diagrams~\cite{9806174}.
For each of these volumes we found, by empirical means,
a {\sl closed--form} evaluation in terms of a few Clausen functions,
evaluated at rational multiples of $\pi$.
Then, thanks to generous advice from Don Zagier, we came to appreciate
that these results form part of a much larger scaffold,
whose unifying feature is the existence of
rational relations between hyperbolic volumes and
Dedekind zeta values. While the existence of such relations is
understood~\cite{DZ:Invent,DZ:JMPS,DZ:Texel}, their precise forms
appear to be unpredictable, thus far, by deductive mathematics.
They are therefore ripe for the application of experimental mathematics.
The work reported herein has involved  a -- to the authors -- quite
intoxicating mix of tools, allowing conjectures to be tested against
known data and rapidly confirmed or rejected as the research progressed.
We will return to these methodological issues in the conclusion.

For any single-complex-place field $K$,
with discriminant $D$ and degree $n$,
we define
\begin{equation}
Z_K:=\frac{(-D)^{3/2}}{(2\pi)^{2n-4}}\,\frac{\zeta_K(2)}{2\zeta(2)}
\label{ZK}
\end{equation}
where the Dedekind zeta value $\zeta_K(2)$
is the sum of the inverse squares of the norms of the
ideals~\cite{Cohen,Zass} of $K$,
and $\zeta(2):=\sum_{n>0}1/n^2=\pi^2/6$
is the corresponding Riemann zeta value. It follows
from~\cite{DZ:Invent} that $Z_K$ is reducible, with
unspecified\footnote{Ratios of coefficients may be specified;
here we ask for values.} rational coefficients,
to Bloch--Wigner dilogarithms $\{D(z_k)\mid z_k\in K\}$.
The volume of a hyperbolic
manifold, for which the single-complex-place field $K$ is
the invariant trace field~\cite{CGHN}, is
systematically~\cite{AHW}
reducible to such dilogarithms and, moreover, is expected
to be some unspecified rational multiple of the very specific
construct~(\ref{ZK}).
Accordingly, we seek coprime integers $a$ and $b$ such that
\begin{equation}
\frac{a}{b}\,{\rm vol}({\cal M})=Z_K\label{ab}
\end{equation}
for a manifold ${\cal M}$ with a single-complex-place invariant
trace field $K$. We call $a/b$ the rational Dedekind-zeta invariant
of ${\cal M}$. It evaluates to unity for manifold ${\rm m}003(-3,1)$,
of conjecturally smallest volume.
It will be seen to take some remarkable values.

In Section~2, we report findings of the rational invariant
for 998 closed manifolds.

In Section~3, we find a further 12 single-complex-place fields
from cusped manifolds, and compute corresponding rational invariants.
We also identify 91 census manifolds whose volumes are reducible to
{\sl pairs\/} of Dedekind zeta values.

Section~4 concerns links, with complementary
manifolds of a richer structure than those
recorded in the Hildebrand--Weeks~\cite{HiW,CHW} census of
cusped manifolds triangulated by no more than 7 tetrahedra.
In particular we identify a 24-crossing 10-component link,
triangulated by 54 ideal tetrahedra with shapes in
the quadratic field ${\bf Q}(\sqrt{-5})$ and determine
its rational Dedekind-zeta invariant. We call it a platonic link,
since its 10 components mimic the vertices and edges of a tetrahedron.
The remaining 4 platonic links, similarly modelled on perfect
solids, are shown to be quartic, after processing the 950
ideal tetrahedra from their triangulations.

Section~5 concerns connections~\cite{9806174,BKP,BDK,BGK,BK15}
between Feynman diagrams, knots and links.
Recent results~\cite{9806174} from quantum field theory suggested
a connection between Dedekind zeta values and volumes of orthoschemes. We
find 10 quadratic fields that forge such connections and in 8 cases
now know links that have corresponding rational Dedekind-zeta invariants.
The remaining 2 cases are quadratic fields with discriminants
$D=-39$ and $D=-84$; these await the discovery of
corresponding Feynman links. Hyperbolic knots with up to 18 crossings,
from Feynman diagrams with up to 11 loops, are also analyzed.
For crossing numbers greater than 9, the physics
in~\cite{BKP,BDK,BGK,BK15}
is a more fertile source of Dedekind zeta values than an analysis of the
maximally symmetric knots found in~\cite{HTW}.

Section~6 offers some conclusions and suggestions for further study.

\section{Dedekind-zeta invariants of closed manifolds}

We have found rational Dedekind-zeta invariants for 998 manifolds
in the Hodgson--Weeks census, which lists $11,031$ closed orientable
manifolds, with 9,218 distinct volumes, all less
than $6.5$, and with all their geodesics having lengths greater than $0.3$.
These 998 manifolds have 224 distinct volumes, which are rationally related,
via~(\ref{ZK},\ref{ab}), to 140 distinct Dedekind zeta values,
each corresponding to an invariant trace field of at least one census
manifold. Table~1 enumerates the fields by degree.
Tables 2--12 give, for each field $K$, a generating polynomial,
together with its discriminant and the rational
relation of $Z_K$ to the volume of the first manifold in the census with
invariant trace field $K$.
The 224 distinct volumes are given in Tables~13a--f, where
it can be observed that numerators and denominators of the 998
invariants are bounded by $a\le9408656$ and $b\le9$. Our methods were as
follows.

\subsection{Examination of 1,200 single-complex-place fields}

It became clear, from studying~\cite{AMS:JB,AMS:SB,AMS:RK,AMS:DZ},
that there are unspecified rational relations of the form~(\ref{ab})
between Dedekind zeta values and volumes of some, but not all,
of the manifolds in the Hodgson--Weeks census. We soon found that
30 of the first 32 census manifolds have rational Dedekind-zeta
invariants, with $a\le46$ and $b\le3$. The 2 exceptions are
manifolds ${\rm m}003(-4,1)$ and ${\rm m}004(+7,1)$,
whose invariant trace fields have 2 complex places.

The next step was to obtain systematic listings of fields with precisely
one complex place. Thanks to the {\tt numberfields}
directory\footnote{{\tt ftp://megrez.math.u-bordeaux.fr/pub/numberfields}}
at the University of Bordeaux, we found
files that order single-complex-place fields, for each degree $n\le7$,
by the magnitudes, $-D$, of the discriminant.
We selected the first 200 fields
for each degree $n\in\{2,3,4,5,6,7\}$ and used the {\tt zetak}
command of Pari to compute 19 digits of the Dedekind zeta values,
$\zeta_K(2)$, of these 1,200 fields. Forming $Z_K$, defined in~(\ref{ZK}),
we made $1200\times11031=13,237,200$ indiscriminate comparisons
between these target fields and the census entries.
This took 19 seconds on a 233MHz Pentium.
We found 842 manifolds, with 175 distinct volumes,
rationally related to 96 distinct Dedekind zeta values, with
numerators and denominators of the rational invariants
bounded by $a\le46$ and $b\le5$.

Since the census provided 17-digit volumes, the probability of any of these
simple rational results being spurious is comfortably less than $10^{-12}$.
Conversely, one needs only 5-digit precision for the volume
to discover the rational Dedekind-zeta invariant in these 842 cases,
and hence to obtain an accurate volume from~(\ref{ZK}).
Only for degrees $n\ge8$
do precision, core memory and CPUtime become issues.
In the $n=12$ case of Table~12, with $|D|=12,476,239,474,594,496$,
these issues require close attention.

The 96 fields that were caught by this 19-second trawl
comprise all those of Tables~2 and~3 and all those
above the lines drawn in Tables~4--7.
Below the lines, and in Tables~8--12,
we took assistance from Melbourne~\cite{CGHN}.

\subsection{Examination of 44 fields found by Snap}

For the remaining $140-96=44$ results of Tables~2--12, we made reference
to the impressive body of files\footnote{{\tt
http://www.ms.unimelb.edu.au/$\,\widetilde{}\,$snap}}
at the University of Melbourne. These were obtained
by Coulson, Goodman, Hodgson and Neumann~\cite{CGHN} (hereafter referred to
as CGHN) using 50-digit precision, in marked distinction
to the 5 digits which suffice for the findings above.

CGHN make no reference to Dedekind zeta values; they do, however, find
single-complex-place invariant trace fields for about 9\% of the
Hodgson--Weeks manifolds. The next step was to examine the
overlap between our single-complex-place fields and theirs.
We found that the CGHN file {\tt closed\_census\_algebras}
contains 95 of our 96 single-complex-place fields.
The exception was the quintic field with $D=-14103$,
for which our low-precision method had
readily yielded the rational invariant $a/b=2$
of the manifold ${\rm s}784(+5,2)$, which was absent from the file
{\tt closed\_census\_algebras}, obtained by CGHN from 50-digit searches
for fields of degree\footnote{We later ran Snap at 100-digit precision,
and found a group field of degree 20.} not exceeding 16.
We then referred to the file {\tt closed.fields}, where
${\rm s}784(+5,2)$ indeed appears, with the expected
invariant trace field, $x^5-2x^4+2x^3-x^2-2x-1$.

Thus assured that all our 96 finds were genuine, we extracted
all of the invariant trace fields from the CGHN files
{\tt closed\_census\_algebras} and {\tt closed.fields},
and asked Pari to determine their signatures and discriminants.
This revealed 44 cases beyond the ranges of our search, namely the
32 below the lines in Tables~4--7 and the 12 in Tables~8--12.
Within our selected ranges, CGHN had found all and only our
96 single-complex-place fields,
indicating the reliability of both their methods and ours.

The next step was obvious, yet computationally demanding: to
determine rational Dedekind-zeta invariants for the
manifolds associated by Snap to the remaining
44 large-discriminant single-complex-place fields.
The {\tt zetak} command of Pari was entirely
adequate to complete Tables~4--7, at 19-digit precision,
yielding numerators and denominators bounded by $a\le976$
and $b\le5$, for which merely 7-digit precision would have
been amply sufficient. The 12 fields of Tables~8--12
presented a much tougher computational challenge.

At degree $n=8$, Pari-GP 2.0.11 began to falter, with {\tt zetak}
yielding results drastically below the requested precision
(as warned in the Pari manual) or running out of memory
(when allocated 100MB of core).
Accordingly we devised a method based on the {\tt dirzetak}
command, which returns the multiplicities, $m_K(n)$, of the
ideals of $K$ with norm $n$, up to some requested maximum, $N$,
that may comfortably extend to $N=10^6$. These multiplicities
are the numerators of the Dirichlet series
\begin{equation}
\zeta_K(2)=\sum_{n>0}\frac{m_K(n)}{n^2}\label{dirz}
\end{equation}
We then computed the truncated sums
$S_N:=\sum_{n=1}^N m_K(n)/n^2$, and accelerated their convergence
by forming $T_{N,M}:=(S_N N-S_M M)/(N-M)$, with $M\approx\frac12N$.
This removes a predictable $O(1/N)$ truncation error,
leaving an $O(1/N^{3/2})$ ``random-walk'' error.
With $N\le10^6$, we were able to achieve
10-digit precision, by averaging out fluctuations in $T_{N,M}$.
Even in the most demanding case of Table~12, where the rational
invariant is a 7-digit integer, we were left with 3 vanishing decimal
places for $a/b=9408656.000\pm0.001$.

Hence we completed the task for Hodgson--Weeks manifolds.
Tables~2--12 give $140$ single-complex-place
invariant trace fields of closed census
manifolds, of which 96 were detected by us without reference to CGHN.
The 224 distinct census volumes rationally related to the Dedekind zeta
values of these fields are given in Tables~13a-f. These 224 distinct volumes
correspond to 998 of the 11,031 entries of the Hodgson--Weeks
census file {\tt ClosedManifolds} available\footnote{The 1995 file {\tt
ftp://ftp.geom.umn.edu/pub/software/snappea/tables/ClosedManifolds}
contains 11,031 entries; unfortunately some of the Dehn fillings
do not correspond to those packaged, internally, with SnapPea.
We enlisted Snap, to check all the surgeries given in Tables~2--13.}
from the Geometry Center at the University of Minnesota.

\subsection{A special case: manifold ${\rm v}3066(-1,2)$}

We note a peculiarity of manifold
${\rm v}3066(-1,2)$, with the same volume as ${\rm v}3066(+1,2)$.
The latter has unit Dedekind-zeta invariant and hence a volume
\begin{equation}
Z_{K_+}:=\frac{59^{3/2}}{(2\pi)^2}\,\frac{\zeta_{K_+}(2)}{2\zeta(2)}
=5.137941201873417769841348339\ldots
\label{v3066}
\end{equation}
where the cubic invariant trace field of ${\rm v}3066(+1,2)$ is
\begin{equation}
K_+:=x^3+2x-1\label{K3p}
\end{equation}
with discriminant $-59$. Snap, working at 100-digit precision,
gave the invariant trace field
of ${\rm v}3066(-1,2)$ as a join of~(\ref{K3p})
with ${\bf Q}(\sqrt{-59})$, generated by a root of the sextic
\begin{equation}
K_-:=x^6-3x^5+10x^4-15x^3+21x^2-14x+4\label{K6m}
\end{equation}
with discriminant $(-59)^3$ and 3 complex places. The Borel
regulator~\cite{CGHN,NY} of ${\rm v}3066(-1,2)$ was given as
$\left[Z_{K_+},-\frac12Z_{K_+},-\frac12Z_{K_+}\right]$.
It thus appears that a single-complex-place
invariant trace field is sufficient, yet {\sl not} necessary,
to give a rational Dedekind-zeta invariant.

\subsection{How may one {\sl derive\/} the rational invariant?}

Some of our results contain non-trivial factors,
with 4-digit and 3-digit primes in
$3\times1223=3669$ and
$2^4\times3\times5\times967=232,080$, from Table~10, and
$2\times3\times7\times11\times149=68,838$, from Table~11.
Their origins are obscure.
The duodecadic example of Table~12 entails
$a/b=2^4\times23\times37\times691=9,408,656$.
We note, though cannot explain, the circumstance
that its largest prime factor also occurs in $\zeta(12)/\pi^{12}
=\frac{691}{638512875}$.
Reverse engineering gave 50 decimal places of
\begin{eqnarray}
\zeta_{K_{12}}(2)&=&
1.06095699592540035751698213238632531266926282705990
\ldots\label{50dp}\\
K_{12}&:=&
x^{12}-3x^{11}-8x^{10}+17x^9+27x^8-19x^7-50x^6
\nonumber\\&&{}
-24x^5+44x^4+37x^3-5x^2-8x-1
\label{K12}
\end{eqnarray}
from high-precision triangulation of
manifold ${\rm v}2824(+4,1)$. We know of no easy way to check~(\ref{50dp})
beyond the first 10 digits, confirmed by accelerated convergence
of a million truncations of~(\ref{dirz}).

Arguments from $K$-theory~\cite{AMS:JB,AMS:DZ} appear powerless
to derive values for the Dedekind-zeta invariant, $a/b$,
though they imply~\cite{DZ:Invent,DZ:JMPS,DZ:Texel}
its rationality, for every manifold whose invariant trace field
has a single complex place.
It thus remains a challenge to derive the simple value $a/b=4$,
for the quadratic manifold ${\rm m}036(-4,3)$, with invariant trace field
${\bf Q}(\sqrt{-7})$ and volume observed, at 1800-digit precision,
to coincide with
\begin{eqnarray}
\frac14Z_{{\bf Q}(\sqrt{-7})}&=&
\frac74\left\{{\rm Cl}_2\left(\frac{2\pi}{7}\right)
+{\rm Cl}_2\left(\frac{4\pi}{7}\right)
-{\rm Cl}_2\left(\frac{6\pi}{7}\right)\right\}\label{Q7}\\
&=&
2.66674478344905979079671246261065004409838388855263
\ldots\label{V7}
\end{eqnarray}
whose reduction to
\begin{equation}
{\rm Cl}_2(\theta):=\sum_{n>0}\frac{\sin(n\theta)}{n^2}\label{Cl}
\end{equation}
is derived in Section~4.
Triangulation gives
\begin{eqnarray}
{\rm vol}({\rm m}036(-4,3))&=&
2D\left(-e^{-i\theta_7}\right)
+D\left(-\mbox{$\frac12$}e^{-i\theta_7}\right)\label{T7}\\
\theta_7&:=&2\arctan\sqrt7
\end{eqnarray}
where the Bloch--Wigner dilogarithm
\begin{equation}
D(z):=\Im{\rm Li}_2(z)+\log|z|\Im\log(1-z)
=\Im\sum_{n>0}\left(\frac{1}{n^2}-\frac{\log|z|}{n}\right)z^n
\label{BW}
\end{equation}
gives the volume of an ideal tetrahedron whose essential dihedral angles
are the arguments of $\{z,1/(1-z),1-1/z\}$. The 1800-digit agreement
of~(\ref{T7}) with~(\ref{Q7}) is compelling evidence that,
with $\theta_7:=2\arctan\sqrt7$,
\begin{equation}
\frac32{\rm Cl}_2(\theta_7)
-\frac32{\rm Cl}_2(2\theta_7)+\frac12{\rm Cl}_2(3\theta_7)=
\frac74\left\{{\rm Cl}_2\left(\frac{2\pi}{7}\right)
+{\rm Cl}_2\left(\frac{4\pi}{7}\right)
-{\rm Cl}_2\left(\frac{6\pi}{7}\right)\right\}\label{todo}
\end{equation}

An elementary proof of this
remarkable relation between Clausen values has escaped both us, and also, it
appears, the authors of~\cite{AMS:JB,AMS:SB,AMS:RK,AMS:DZ}.
To cast it in a less classical
form, we multiply~(\ref{todo}) by 2, and transform to Bloch--Wigner
functions, obtaining
\begin{equation}
4\,D\left(\frac{3+i\sqrt7}{4}\right)+
2\,D\left(\frac{3+i\sqrt7}{8}\right)=
7\,D\left(2\frac{1+i\tan{\pi\over7}}
{3-\tan^2{\pi\over7}}\right)
\label{modern}
\end{equation}
where the argument on the r.h.s.\ is a root of the totally complex
sextic
\begin{equation}
K_6:=x^6-4x^5+9x^4-8x^3+4x^2-2x+1\label{K6}
\end{equation}
with discriminant $(-7)^5$. The l.h.s.\ of~(\ref{modern}) is the
volume of the cusped manifold ${\rm s}776$, which is
complementary to the 3-component 6-crossing link
$6^3_1$. Its triangulation
immediately yields 6 ideal tetrahedra, with 2 distinct shapes in the
quadratic field ${\bf Q}(\sqrt{-7})$.
The challenge is to prove that by combining
these 6 one forms the same volume as from the 7 congruent sextic shapes
on the r.h.s.\ of~(\ref{modern}).
This is a modern version of our puzzle~(\ref{todo}) in classical analysis:
how may one relate Clausen values at multiples of
$\theta_7:=2\arctan\sqrt7$ to Clausen values at multiples of $2\pi/7$,
when there appears to be no non-trivial relation between
trigonometric functions of the two sets of angles?
A far greater challenge would be to derive, rather than merely measure,
the integer Dedekind-zeta invariant $2^4\times23\times37\times691$
of the closed manifold ${\rm v}2824(+4,1)$.

\section{Cusped manifolds and join fields}

Now we turn to the study of cusped manifolds,
complementary to knots and links.
Packaged with SnapPea~\cite{SnapPea}, there
are the 4,815 orientable cusped manifolds of
the Hildebrand--Weeks census~\cite{HiW},
triangulated by no more than 7 ideal tetrahedra,
and 114 non-orientable manifolds, triangulated by no more than 5.
Making $4929\times1200$ comparisons
with the Dedekind zeta values of Section~2.1, we found 7
new single-complex-place fields, beyond those from
the Hodgson--Weeks census. Then the CGHN file {\tt cusped.fields}
confirmed these finds and yielded the remaining 5 fields of Table~14,
with larger discriminants. In Table~14, we give the rational
Dedekind-zeta invariant of a selected cusped manifold for each of the
12 new fields.

Our tally of single-complex-place fields
is now $96+44+7+5=152$.
Among the $4,929$ cusped manifolds, we found
312 whose volumes are rationally related to one of
the 152 Dedekind zeta values, with numerators and denominators
in~(\ref{ab}) bounded by $a\le2853$ and $b\le7$.

We also sought integer relations of the form
\begin{equation}
a\,{\rm vol}({\cal M})=b_1Z_{K_1}+b_2Z_{K_2}
\label{ab12}
\end{equation}
corresponding to a cusped or closed
manifold ${\cal M}$ whose invariant trace field
is the join of single-complex-place fields $K_1$ and $K_2$, or is
a subfield of this join. Restricting $K_1$ and $K_2$ to the 152 fields of
Tables~2--12 and Table~14, we found 91 census manifolds, with 26
distinct volumes, rationally related, via~(\ref{ab12}),
to pairs of Dedekind zeta values.
Details of the 29 cusped and 62 closed manifolds
are provided by Tables~15--18. Two of the 26 integer relations
entail join fields noted in~\cite{CGHN}, namely the first in
Table~16 and the second in Table~17.
We used Pari's {\tt nfisincl} command to confirm
that all 6 of the quartic invariant trace fields in
Table~18 are subfields of the octadic joins.
In 4 of these 6 cases, distinct values of
$b_1/b_2$ occur, for the same invariant trace field.

We believe that we have exhausted the 3-term relations between
census volumes and pairs of the
152 target Dedekind zeta values.
David Bailey's impressively efficient,
arbitrary precision~\cite{TRANSMP},
implementation of PSLQ~\cite{PSLQ}
found the 91 relations in 40 minutes and then took 17 hours to
exhaust the
$(4815+114+11031)\times{152\choose2}=183,156,960$ indiscriminate
possibilities. The PSLQ search rate was thus a healthy 3 kHz.
No reduction to 3 distinct Dedekind zeta values was detected,
presumably because the census volumes are kept
deliberately small.

\section{Links whose volumes link Clausen values}

Our order of presentation is the reverse of the order
in which we obtained results. Our primary
motivation was to elucidate connections between Clausen values,
revealed by studies of Feynman diagrams~\cite{9806174},
on which we comment in Section~5.

This section concerns links whose rational Dedekind-zeta
invariants link Clausen values. An example is provided by the
attractive alternating\footnote{The viewer is asked
to alternate all the crossings in Fig.~1 and Figs.~3--8.}
link of Fig.~1, discovered
as a result of work on light-by-light scattering,
reported in Section~5.
A second is provided by the
non-alternating\footnote{The viewer is asked
to supply the uniquely non-alternating non-trivial crossings in Fig.~2.}
daisy-chain link of Fig.~2.
These form part of our study of Dedekind-zeta invariants
of quadratic links.

Next in chronological sequence, came the
higher-degree Dedekind-zeta invariants of cusped manifolds, in Section~3.
Finally, we undertook the systematic study of closed manifolds,
culminating with the 12th-degree Dedekind-zeta invariant of Section~2.
Like most stories, it benefits from signposting.

\subsection{Signpost: integer sequence A003657}

Mathematically speaking, the results of Section~2, while numerically
striking, are of a rational character~\cite{DZ:Invent,DZ:JMPS,DZ:Texel}
that was expected by specialists of $K$-theory
(which we are not).
The challenge is not to understand why a rational Dedekind-zeta
invariant exists, but to learn how to derive (as opposed to measure) it.
The cusped results of Section~3 were of the same character as for
closed manifolds:
hyperbolic knots and links have rational Dedekind-zeta invariants
if their complementary cusped manifolds have single-complex-place invariant
trace fields.  Here, in Section~4, we address the question
raised by the sparsity of quadratic entries in Tables~2 and~14: where are
the quadratic links with (negated) discriminants beyond the
first 4 entries in the integer sequence A003657 of Neil
Sloane's on-line encyclopedia\footnote{{\tt
http://www.research.att.com/$\,\widetilde{}\,$njas/sequences}}
\begin{equation}
\underline3,\underline4,\underline7,\underline8,
\underline{11},\underline{15},19,\underline{20},23,\underline{24},31,35,
\underline{39},40,43,47,51,52,55,56,59,67,68,71,79,83,
\underline{84}\ldots\label{iq}
\end{equation}
Only the first 4 cases figure in the Hildebrand--Weeks census.
Here we identify 4 more underlined cases,
with $-D\in\{11,15,20,24\}$, where the platonic
link of Fig.~1 furnishes an example with $D=-20$, and the
daisy-chain link of Fig.~2 furnishes an example with $D=-24$.
Then, in Section~5, we shall explain how we were led to Figs.~1 and~2,
by consideration of the physical process of light-by-light scattering,
and why its Feynman diagrams suggest that the remaining 2 underlined
cases, with $D=-39$ and $D=-84$, will be of similar symmetrical appeal,
and similar analytical mystery.

\subsection{Dirichlet character}

The simplest example of a number field with a single complex place
is an imaginary-quadratic field ${\bf Q}(\sqrt{-d})$, where $d$
is a square-free positive integer. When $d\equiv3$ (mod 4), the
discriminant is $D=-d$; otherwise it is $D=-4d$.

It is proven in~\cite{Berndt} that
\begin{equation}
\frac{\zeta_K(2)}{\zeta(2)}=\sum_{n>0}\frac{\chi_K(n)}{n^2}\label{chi}
\end{equation}
where $\chi_K$ is the real Dirichlet character of the group
of units of the field ${\bf Z}/|D|$ and $D$ is the discriminant
of the imaginary-quadratic field $K$.
The Dirichlet character is related to the Jacobi (or Kronecker)
symbol by
\begin{equation}
\chi_K(n)=\left(\frac{D}{n}\right)\label{Jacobi}
\end{equation}
which vanishes if ${\rm gcd}(D,n)>1$. When $D$ is odd, one may
use the alternative form $\left(\frac{n}{|D|}\right)$.

Using~(\ref{chi}) in~(\ref{ZK}), at $n=2$, one
happily disposes of powers of $\pi$.
We dispose of $\sqrt{-D}$, by using the imaginary-quadratic
result~\cite{Berndt}
\begin{equation}
\sum_{n>0}\frac{\chi_K(n)}{n^2}=
\frac{1}{\sqrt{-D}}\sum_{-D>k>0}\chi_K(k)
{\rm Cl}_2\left(\frac{2\pi k}{|D|}\right)\label{chik}
\end{equation}
which yields a finite sum over the elements of the group of units.
Finally, we dispose of a factor 2, via the reflection relations
$\chi_K(k)=-\chi_K(|D|-k)$ and
${\rm Cl}_2(\theta)=-{\rm Cl}_2(2\pi-\theta)$, to obtain the
readily computable result
\begin{equation}
Z_K=Z_{|D|}:=D\sum_{-D>2k>0}\left(\frac{D}{k}\right)
{\rm Cl}_2\left(\frac{2\pi k}{D}\right)\label{ZKq}
\end{equation}
with the magnitude of the discriminant identifying
the imaginary-quadratic field.

Thanks to David Bailey~\cite{TRANSMP}, it is now a routine
matter to evaluate~(\ref{ZKq}) to high precision\footnote{{\tt
http://science.nas.nasa.gov/Groups/AAA/db.webpage/mpdist/mpdist.html}},
using~\cite{Lewin}
\begin{equation}
\frac{{\rm Cl}_2(\theta)}{\theta}
=1-\df12\log(\theta^2) +\sum_{n=1}^\infty\frac{\zeta(2n)}{2n^2+n}
\left(\frac{\theta}{2\pi}\right)^{2n}
\label{cl2}
\end{equation}
which converges faster than $1/9^n$, for
$\theta\in[0,2\pi/3]$. An angle in $[2\pi/3,\pi]$ may be transformed
to a pair in $[0,2\pi/3]$, by rearrangement of the duplication formula
\begin{equation}
\df12{\rm Cl}_2(2\theta)={\rm Cl}_2(\theta)-{\rm Cl}_2(\pi-\theta)
\label{dub}
\end{equation}
Rather than compute $\zeta(2n)/(2\pi)^{2n}
=\frac12|B_{2n}|/(2n)!$ recursively, from the Bernoulli numbers,
we used a FFT program written by David Bailey to evaluate
$\{\zeta(2n)\mid 0\le n<2^{m}\}$, in one fell swoop,
by a multi-dimensional generalization of Newton's method.
With $m=11$, the initial outlay, to tabulate 2048 coefficients
to 1800 digits, was 8 minutes; then 1800 good digits
of any Clausen value in $\{{\rm Cl}_2(\theta)\mid 0<\theta<2\pi/3\}$
are obtainable in less than 2 seconds on a 533MHz machine.
Low-precision results for $Z_{|D|}$ are given in Table~19.

It follows from~(\ref{ZKq})
that every discovery of a rational Dedekind-zeta invariant
for a quadratic link gives a relation between a
set of Clausen values at angles whose tangents involve $\sqrt d$, and a set
of Clausen values at angles which are multiples of $2\pi/|D|$.
We have already seen this,
rather strikingly,
in the case $-D=d=7$ of~(\ref{todo}), where the hyperbolic
6-crossing 3-component
link\footnote{We use the notation of~\cite{Rolfsen}, whose appendix
give drawings of links up to 9 crossings.} $6^3_1$ mediates between
Clausen values, and hence the Bloch--Wigner dilogarithms
of~(\ref{modern}), with very different types of argument
on the left and right.

We now address the question:
which links have volumes that link Clausen values?

\subsection{Figure-8 knot at $D=-3$}

The first case, at $D=-3$, is elementary. The
unique\footnote{See Helaman Ferguson's sculpture at {\tt
http://www.mtholyoke.edu/acad/math/ma/sculpture.htm}}
arithmetic knot is the figure-8 knot of Fig.~3. Its volume
is the first entry of Table~19.
It links Clausen values at
$2\arctan\sqrt3$ and $2\pi/3$. This relation clearly
poses no analytical puzzle, since the angles are identical.

\subsection{Whitehead link at $D=-4$}
Just as the simplest hyperbolic knot, $4_1$, gave the answer at $D=-3$,
so the simplest hyperbolic 2-component link, namely the
Whitehead link $5^2_1$ of Fig.~4, gives the answer at $D=-4$.
Its volume is the second entry of Table~19.
It links Clausen values at
$2\arctan\sqrt1$ and $2\pi/4$, which are again identical.

\subsection{Link $6^3_1$ at $D=-7$}
The Dedekind-zeta invariant of the link $6^3_1$,
illustrated in Fig.~5, is measured to be $2$, corresponding to
\begin{equation}
2\left\{3{\rm Cl}_2(\theta_7)
-3{\rm Cl}_2(2\theta_7)+{\rm Cl}_2(3\theta_7)\right\}=Z_7=
7\left\{{\rm Cl}_2\left(\frac{2\pi}{7}\right)
+{\rm Cl}_2\left(\frac{4\pi}{7}\right)
-{\rm Cl}_2\left(\frac{6\pi}{7}\right)\right\}\label{d7}
\end{equation}
with $\theta_7:=2\arctan\sqrt7$. At 1800-digit precision,
PSLQ finds this to be the sole integer relation between
the 6 Clausen values. The combination
on the left is selected by hyperbolic geometry; that
on the right by the Dirichlet character.
Thus the rational Dedekind-zeta invariant of link $6^3_1$
encodes a highly non-trivial analytical relation between Clausen values.
A derivation of $a/b=2$ would prove~(\ref{d7}).
Conversely, a proof of~(\ref{d7}) would derive $a/b=2$.
Where may one find {\sl any} such proof?

\subsection{Link $9^2_{40}$ at $D=-8$}
As exemplar of the next non-trivial relation between Clausen
values, we select the 9-crossing 2-component link
$9_{40}^2:=(\sigma_1^2\sigma_2^{-1})^3$ of Fig.~6.
Its Dedekind-zeta invariant is observed to be unity,
which implies that its volume is equal to each side of
\begin{equation}
\df12\left\{27{\rm Cl}_2(\theta_2)-9{\rm Cl}_2(2\theta_2)
+{\rm Cl}_2(3\theta_2)\right\}
=Z_8=8\left\{{\rm Cl}_2\left(\frac{2\pi}{8}\right)
+{\rm Cl}_2\left(\frac{6\pi}{8}\right)\right\}
\label{d8}
\end{equation}
with $\theta_2:=2\arctan\sqrt2$. As in the case of~(\ref{d7}),
a proof is lacking.

\subsection{A 12-crossing 4-component link at $D=-11$}

In~\cite{CGHN} it was noted that neither the closed nor the cusped census
entails the field ${\bf Q}(\sqrt{-11})$. In~\cite{9806174}
it was observed that
the 12-crossing 4-component link of Fig.~7, with braid word
$(\sigma_1^{}\sigma_2^{-2}\sigma_3^{}\sigma_2^{-2})^2$,
entails this field. Its Dedekind-zeta invariant is
observed to be unity, corresponding
to the relation
\begin{equation}
15{\rm Cl}_2(\theta_{11})-10{\rm Cl}_2(2\theta_{11})
+{\rm Cl}_2(5\theta_{11})=Z_{11}=11\sum_{k=1}^5\left(\frac{k}{11}\right)
{\rm Cl}_2\left(\frac{2\pi k}{11}\right)
\label{d11}
\end{equation}
where $\theta_{11}:=2\arctan\sqrt{11}$ and $\left(\frac{k}{11}\right)$
is the Jacobi (or Legendre) symbol for the Dirichlet character.
Again, we lack a proof.

\subsection{A 12-crossing 3-component link at $D=-15$}

The next case likewise comes from~\cite{9806174}.
The 12-crossing 3-component link of Fig.~8, with braid word
$(\sigma_1^2\sigma_2^{-2})^3$,
has Dedekind-zeta invariant empirically equal to 2,
at 1800-digit precision. This corresponds to the relation
\begin{equation}
24{\rm Cl}_2(\theta_{5,3})-12{\rm Cl}_2(2\theta_{5,3})
-8{\rm Cl}_2(3\theta_{5,3})
+6{\rm Cl}_2(4\theta_{5,3})=Z_{15}
=15\sum_{k=1}^7\left(\frac{k}{15}\right)
{\rm Cl}_2\left(\frac{2\pi k}{15}\right)
\label{d15}
\end{equation}
with $\theta_{5,3}:=2\arctan\sqrt{5/3}$. Again, where is the proof?

\subsection{A self-dual platonic link at $D=-20$}

The reader might now expect us to consider the case $D=-19$.
For reasons that will be given in Section~5, we skip to
$D=-20$, where we were rewarded by the
splendidly symmetric alternating
platonic link drawn
in Fig.~1.
Its Dedekind-zeta invariant was measured to be unity.
This innocent-sounding statement translates to our 5th unproven relation
\begin{equation}
36{\rm Cl}_2(\theta_5)-30{\rm Cl}_2(2\theta_5)
+4{\rm Cl}_2(3\theta_5)+3{\rm Cl}_2(4\theta_5)=Z_{20}
=20\sum_{k\in\{1,3,7,9\}}{\rm Cl}_2\left(\frac{2\pi k}{20}\right)
\label{d20}
\end{equation}
with $\theta_5:=2\arctan\sqrt5$. Between these 8 Clausen values,
PSLQ found no other relation.

We found that, far from complicating the result, the 4 Clausen values
on the left simplify the 54 Bloch--Wigner values of the triangulation
in ${\bf Q}(\sqrt{-5})$.
There is, of course, an infinity of rewritings of~(\ref{d20}),
obtained by adding, on the left, combinations
of Bloch--Wigner values that algebraically~\cite{AMS:SB} sum to zero,
by virtue of special cases of the classical~\cite{Lewin}
2-variable 5-dilogarithm relation
of Abel, which is easily proved by differentiation.
Here, as elsewhere, we expose the classical analysis
that remains to be done.

\subsection{A 48-crossing daisy-chain link at $D=-24$}

The attentive reader will now expect us to skip the case $D=-23$
and jump to the 8th underlining in~(\ref{iq}), at $D=-24$.
This is precisely what we did, though the
character of the result was not what we
first supposed. It seemed to us that the hallmark of past success,
in finding links that link Clausen values, was to achieve
the largest possible symmetry group. What, we asked ourselves,
could pack a better symmetry-to-crossing ratio
than the remaining 4 platonic alternating links?

Let us denote the tetrahedral link of Fig.~1 by $T:=24^{10}_{\rm tet}$.
The 4 non-self-dual alternating platonic links have
components that mimic the vertices and edges of the
cube $C:=48^{20}_{\rm cub}$,
octahedron $O:=48^{18}_{\rm oct}$,
dodecahedron $D:=120^{50}_{\rm dod}$, and
icosahedron $I:=120^{42}_{\rm ico}$,
where, for example, the last has 120 alternating crossings
of its 42-components, which mimic the 12 vertices and 30
edges of the perfect solid with 20 faces.
These links are highly symmetrical, yet none of their 4 volumes
yielded the sought-for rational relation
to a quadratic Dedekind zeta value. Later, we show that they
are quartic links.

Instead, we found an answer to our 8th question by
a more child-like construction: a non-alternating daisy chain.
By a daisy chain we mean a link
each of whose components has 2 crossings with a neighbour
on one side, and 2 crossings with a neighbour on the other,
with the whole forming a circle, as in Fig.~2.
By a non-alternating daisy chain,
we mean one in which the 4 crossings of each component occur in an order
over-over-under-under, while still linking with neighbours.
A little doodling
should convince the reader that non-alternating daisy-chain links
must have an even number of components.

A delightful feature of Jeff Weeks' program SnapPea~\cite{SnapPea}
is that it enables
one to draw such chains quickly, and then ask whether the resultant
non-alternating link has quadratic shapes in its triangulation.
Using A.C.~Manoharan's
port of SnapPea to Windows95\footnote{{\tt
http://home.att.net/$\,\widetilde{}\,$Manoharan/SnapPea/snappea.html}},
we inferred, from instances with up to 96 crossings,
that the non-alternating daisy-chain link with $2n\ge6$ components,
and hence $4n\ge12$ crossings,
has hyperbolic volume
\begin{equation}
V_{2n}=8nD\left(i\tan\frac{(n-2)\pi}{4n}\right)\label{daisy}
\end{equation}
The first 4 hyperbolic non-alternating daisy chains yield
results with $-D\in\{3,4,8,20\}$:
\begin{eqnarray}
V_6&=&4Z_4\label{dc6}\\
V_8&=&2Z_8\label{dc8}\\
V_{10}&=&\df12Z_{20}+V_6\label{dc10}\\
V_{12}&=&20Z_3\label{dc12}
\end{eqnarray}
Then nothing interesting happens until we reach 24
components, with 48 crossings, where
\begin{equation}
V_{24}=Z_{24}+V_{8}\label{dc24}
\end{equation}
hits the target, at $D=-24$.

SnapPea illustrates this nicely. Drawing daisy chains with
6,~8,~10,~12,~24 components, and inspecting the triangulations,
one detects the square roots of 1,~2,~5,~3,~6. The last,
illustrated in Fig.~2,
delivers a result for $Z_{24}=V_{24}-V_{8}$, namely
\begin{equation}
60{\rm Cl}_2(\theta_{3,2})-18{\rm Cl}_2(2\theta_{3,2})
-4{\rm Cl}_2(3\theta_{3,2})+3{\rm Cl}_2(4\theta_{3,2})=Z_{24}
=24\sum_{k\in\{1,5,7,11\}}{\rm Cl}_2\left(\frac{2\pi k}{24}\right)
\label{d24}
\end{equation}
with $\theta_{3,2}:=2\arctan\sqrt{3/2}$.
So, for the 6th time, we have a relation that is as easy
to check numerically as it appears hard to derive.

\subsection{Quartic platonic links}

Our first guess, that the 4 non-self-dual platonic links might
yield quadratic number fields, failed. Nonetheless, it is
interesting to determine their number fields,
and hence obtain accurate volumes.
This is clearly a taxing job.
To prepare for it, we first tackled
a problem of similar complexity, where
an accurate answer could be inferred.

We found that by adding
$n$ concentric components at each of the 4 ``vertices'' of
Fig.~1, one obtains a volume $Z_{20}+2nZ_{15}$ for the link
with $24+24n$ crossings and $10+4n$ components,
when $n\le5$.
At $n=6$, SnapPea was given the 34-component link with 168 crossings,
which was triangulated to give 632 ideal tetrahedra, each having a volume
$D(z_k)$ with $z_k\in\mbox{\bf Q}(\sqrt{-5})$ or
$z_k\in\mbox{\bf Q}(\sqrt{-15})$.
We output the 632 shapes, and computed, at 1800-digit precision,
the $3\times632=1,896$ Clausen values entailed by
\begin{eqnarray}
D(z)&=&V(\arg(z),-\arg(1-z))\label{arg}\\
V(\theta_1,\theta_2)&:=&\df12\left\{{\rm Cl}_2(2\theta_1)
+{\rm Cl}_2(2\theta_2)
-{\rm Cl}_2(2\theta_1+2\theta_2)\right\}
\label{3cl}
\end{eqnarray}
obtaining 1800-digit agreement with the expectation
$Z_{20}+12Z_{15}=502.408032\ldots$
for the volume.
This gave us confidence that we could process the 950
ideal tetrahedra entailed by the remaining 4 platonic links.

We obtained from SnapPea the low-precision volumes
\begin{eqnarray}
{\rm vol}(C)\approx{\rm vol}(O)&\approx&{\tt 114.537611}\label{CO}\\
{\rm vol}(D)\approx{\rm vol}(I)&\approx&{\tt 310.913145}\label{DI}
\end{eqnarray}
from $128+130+346+346=950$ ideal tetrahedra.

The duality between cube and octahedron, and between dodecahedron
and icosahedron, is gratifying. Such duality is not restricted to platonic
links. More generally, suppose that we have a link, $L$,
with an alternating projection in which the components may be separated into
two classes: vertex (V) components and edge (E) components, with pairs
of crossings only between V and E components, and every E
connecting a pair of V's. Thus the
crossing number is 4 times the number of E components.
Now shrink the V components to true vertices (some of which may
be divalent) and the E components to true edges that connect these points,
to obtain a
planar graph, $P$. Then construct the dual graph, $P^*$,
whose vertices lie in the regions of the plane partitioned by the edges
of $P$ and whose edges thus cross the edges of $P$. Now thicken $P^*$
to obtain the alternating link $L^*$ dual to $L$.
The crossing numbers of $L$ and $L^*$ are equal,
but the numbers of components need not be.
For example $C:=48^{20}_{\rm cub}$ has more components than
$C^*:=O:=48^{18}_{\rm oct}$, and $D:=120^{50}_{\rm dod}$ has more than
$D^*:=I:=120^{42}_{\rm ico}$.
On the basis of these and further tests,
we conjecture that ${\rm vol}(L)={\rm vol}(L^*)$, for every shrinkable
link $L$.

To identify the number fields of the $\{C,O\}$ and
$\{D,I\}$ pairs, we first examined the cusp shapes,
and found that
\begin{equation}
K_C:=\mbox{\bf Q}\left(\sqrt{8\sqrt{2}-15}\right)\label{FC}
\end{equation}
gave a simple fit to the cusp shapes of $\{C,O\}$,
at the 12-digit precision provided by SnapPea, while
\begin{equation}
K_D:=\mbox{\bf Q}\left(\sqrt{-12\sqrt{5}-31}\right)\label{FD}
\end{equation}
similarly fitted the cusp shapes of $\{D,I\}$.

We then output $z$-values of the triangulations.
Fitting these was not easy, since the data now consisted
of numerical values, to 12 decimal places, of
$1,900$ real or imaginary parts, each of which
was supposed to be fittable by two powers
of $\sqrt{15-8\sqrt{2}}$, in the cubical and octahedral cases,
or $\sqrt{31+12\sqrt{5}}$, in the dodecahedral and icosahedral cases.
In each of the $1,900$ cases we required a significant integer relation
between 3 numbers: the 12-digit SnapPea datum and the two powers,
namely 0 and 2 for a real part, or 1 and 3 for an imaginary part.
Whether this can be done by a reliable and uniformly automated method
depends upon the largest integer. If it has more than 4 digits, the matter
is moot, since fitting 12-digit data with 3 integers with up to 4 digits
is already a parlous business. In fact our first attempt yielded garbage
in a significant fraction of the $1,900$ cases.

Fortunately, 3 features of the mathematics enable us to crack this tough
nut. First, the 6-fold symmetry of the Bloch--Wigner function
\begin{equation}
D(z)=D(1/(1-z))=D(1-1/z)=D(1/\overline{z})
=D(1-\overline{z})=D(\overline{z}/(\overline{z}-1))
\label{6fold}
\end{equation}
gave us 6 bites at each cherry. Secondly, number theory
suggested that we might do well to make the Ansatz
\begin{equation}
z=\frac{1}{I_z}\sum_{p=0}^3 I_p u^p\label{ansatz}
\end{equation}
with $6\times950\times5=28,500$ integers
relating 6 equivalent $z$ values of 950 ideal volumes
to 4 powers of $u=\frac12(1+i\sqrt{15-8\sqrt{2}})$ or
$u=\frac12(1+i\sqrt{31+12\sqrt{5}})$.
Finally, we were prepared for even the simplest of the 6 fits to any
shape to contain the prime factor $97=15^2-2\times8^2$ of the
cubical/octahedral discriminant, or $241=31^2-5\times12^2$ in the
dodecahedral/icosahedral case.

These 3 features allowed us to devise
a diversity of algorithms that yielded, eventually, total fits which
we consider to be corporately indubitable.
Our confidence came from a fourth mathematical feature,
which at first sight appeared to make life difficult, namely
that there was virtually no overlap between problems that were
supposed to be related by duality. In fact we found only two
distinct ideal volumes that contributed to both the
dodecahedron and icosahedron, and each of these was unambiguous.
The absence of further overlap
doubled the computational load, yet made the
resulting numerical agreement of volumes, to 1800 digits, a
potent signal of success. It seems unlikely that
a misidentification of one of the ideal volumes, for the dodecahedron,
could produce the same effect as a misidentification of a quite
different ideal volume, for the icosahedron.

Thus we believe that we have found exact (and also very lengthy) expressions
of the form $\sum_{k=1}^{n}D(z_k)$, with $n=128$ and
$z_k\in\mbox{\bf Q}(\sqrt{8\sqrt{2}-15})$,
for the cubical/octahedral volume of Table~20,
and with $n=346$ and $z_k\in\mbox{\bf Q}(\sqrt{-12\sqrt{5}-31})$,
for the dodecahedral/icosahedral volume of Table~21. Thanks to David
Bailey's evaluations of $\zeta(2n)$, described in Section~4.2,
we were able to obtain $1,800$ good digits of
$2,850$ Clausen values in less than 2 hours.

\section{Knots and links from Feynman diagrams}

Finally, we arrive at the motivating idea for this work:
values of Feynman diagrams.

\subsection{Hyperbolic Feynman links}

In~\cite{DD}, Andrei Davydychev and Bob Delbourgo
made a fine discovery: the dilogarithms of the box diagram for
particle scattering are those which give the volume of a tetrahedron
in hyperbolic 3-space (or its analytic continuation).

In the 3-dimensional~\cite{Nickel}
studies of statistical physics, one-loop Feynman diagrams
yield logarithms;
in the 4 space-time dimensions of relativistic
quantum field theory (QFT), they yield dilogarithms~\cite{tHV}.
A connection between the dilogs of QFT and those of hyperbolic geometry
was considered in~\cite{OW}. The achievement of~\cite{DD} is to
derive a fairly simple relation between the value of
any scalar box diagram, in 4 space-time dimensions,
and the volume of an explicit tetrahedron in a 3-space of
constant curvature.
There is nothing ideal about this tetrahedron: in general it
has 6 essential dihedral angles,
determined by the 10 physical quantities in the problem:
the 4 external masses, the 4 internal masses, and the Mandelstam
variables $s$ and $t$, related to the energy at which the process occurs
and the angle of scatter. Already one sees a nice simplification, with
9 dimensionless ratios of physical quantities collapsing to 6
essential dihedral angles. So far, however, the question of number
theory does not arise, since in the generic physical situation
the kinematic quantities are real variables, and hence no algebraic
number field is implied for the arguments of the dilogarithms.

Now consider the process of light-by-light scattering,
where the external (photon)
masses vanish and there is a common internal (electron) mass,
normalized to unity. There remain
the Mandelstam variables, $s$ and $t$.
Following work that involved comparable Clausen
values~\cite{Sixth,Ising},
it was observed in~\cite{9806174} that light-by-light scattering
yields remarkable results at $s=t=n\in\{4,5,6,7\}$,
where the dilogs give rational multiples of the volumes of links.
These links all figure in Section~4, above,
where their volumes were rationally related to $Z_{|D|}$,
in~(\ref{ZKq}). The $n=4$ case
gives a rational multiple of $Z_4$, corresponding to the Whitehead link;
$n=5$ relates to $Z_{15}$ and the link $(\sigma_1^2\sigma_2^{-2})^3$;
$n=6$ to $Z_3$ and the figure-8 knot; $n=7$
to $Z_7$ and $6^3_1$.
Moreover, a rational multiple of the volume, $Z_8$, of the
link $9_{40}^2:=(\sigma_1^2\sigma_2^{-1})^3$
is obtained for Mandelstam variables $s=\frac12t=4$.

In the course of the present work
we discovered that the hyperbolic
volume, $Z_{20}$, of the self-dual alternating
platonic link of Fig.~1 corresponds to the
values\footnote{For a scattering
above the electron-positron threshold, with $s>4$ and $t<0$,
unitarity makes the amplitude complex.
A relation to a real hyperbolic volume
is obtained by analytic continuation to $t>4$.}
$s=\frac12t=5$ for the Mandelstam variables.
Perhaps not even Delbr\"uck and Meitner\footnote{In 1933,
Max Delbr\"uck (1906--81) and Lise Meitner~(1878--1968) foresaw
non-linear effects in quantum electrodynamics.
See {\tt http://www.nobel.se/laureates/medicine-1969-1-bio.html}
for Delbr\"uck's subsequent work on molecular genetics
and sensory physiology.}
could have imagined that light-by-light scattering
would spawn a tetrahedral hyperbolic link.

\subsection{Dedekind-zeta invariants of Feynman orthoschemes}

Let us try to disentangle this remarkable circumstance from
its physical origin. What is the common characteristic of the
Davydychev--Delbourgo (DD) hyperbolic tetrahedron at
those values of the physical quantities
which gave the volumes of links?

In terms of~(\ref{3cl}), we define the 3-argument dilogarithm
\begin{eqnarray}
S(\psi_1,\psi_2,\psi_3)
&:=&V(\delta+\psi_1,\delta-\psi_1)+V(\delta+\psi_3,\delta-\psi_3)\nonumber\\
&+&V(\df12\pi+\psi_2-\delta,\df12\pi-\psi_2-\delta)
+V(\df12\pi-\delta,\df12\pi-\delta)\label{Schl}
\end{eqnarray}
with an auxiliary angle $\delta\in[0,\pi/2]$ satisfying~\cite{DD}
\begin{equation}
\tan^2\delta
=\frac{\cos^2\psi_2}{\cos^2\psi_1\cos^2\psi_3}
-\tan^2\psi_1\tan^2\psi_3\label{delta}
\end{equation}
Then the Schl\"afli/Lobachevsky~\cite{Lob,Schl,Coxeter,Vinberg}
function~(\ref{Schl})
is 4 times the volume of
a bi-rectangular hyperbolic tetrahedron, with essential
dihedral angles $\psi_1,\psi_2,\psi_3$.
The edges with angles $\psi_1$
and $\psi_3$ are opposite, while that with $\psi_2$ is adjacent to
each. The remaining 3 dihedral angles are right angles.
With 3 essential angles, a bi-rectangular tetrahedron
is an example of an orthoscheme~\cite{AMS:RK}.
With a common internal mass, and a common external mass,
the DD tetrahedron comprises 4 identical orthoschemes,
with~(\ref{Schl}) giving its volume.

The next step was to calculate the essential angles $\psi_k$
that had given a rational relation to the volume of a link.
In all cases we found that $\psi_k\in\{0,\pi/6,\pi/4,\pi/3\}$.
The final step was clear: to compute all such instances of~(\ref{Schl}),
in the hope of finding more Feynman
orthoschemes that are rationally related
to Dedekind zeta values. Taking account of the positivity of~(\ref{delta}),
and the symmetry $S(\psi_1,\psi_2,\psi_3)=S(\psi_3,\psi_2,\psi_1)$,
there are 36 possibilities to consider, of which the physics
in~\cite{9806174}
had already shown 5 to be fruitful. We had a lively expectation
of further rational relations to $Z_{|D|}$. We were thus
totally delighted by the following, totally rational, finding.

\begin{quotation}
\noindent{\bf Discovery:}\quad All real instances of~(\ref{Schl}) with
$\psi_k\in\{0,\pi/6,\pi/4,\pi/3\}$ are rational multiples of
$Z_{|D|}$ with $|D|\in\{3,4,7,8,11,15,20,24,39,84\}$.
Table~22 gives the 36 empirical relations. All the relations are realized
by Feynman diagrams; in the cases with $\psi_2\ne0$, one has merely
to give the external particles a suitable common mass.
\end{quotation}

It was trivial to decide which value of $Z_{|D|}$
to try in each of the 36 cases: one has only
to examine the square root of~(\ref{delta}) to determine $D$.
Had we been less result-oriented
we might have taken time out to
recast\footnote{There is nothing complex, or ideal,
in the physical problem that led to our discoveries:
integration over real Feynman parameters yields the volume of
a tetrahedron, none of whose vertices are at infinity.}
each of the 36 searches in terms of 4 complex arguments
of Bloch--Wigner dilogarithms for the 4 ideal parts of~(\ref{Schl}),
and then used algebra~\cite{AMS:SB,AMS:DZ} to determine, in  advance,
whether a rational (but unspecified!) number would result from
dividing the volume by $Z_{|D|}$. We were content with the faster process
of division, which gives the concrete result. Throughout this work,
the issue is not the existence of rationals, but their values.
The rational numbers of Table~22
in the cases with $-D\ge7$ are at least as hard to derive as
those in~(\ref{d7}--\ref{d20},\ref{d24}). For $D=-39$,
we found that
\begin{equation}
24{\rm Cl}_2(\theta_{13,3})-15{\rm Cl}_2(2\theta_{13,3})
+{\rm Cl}_2(6\theta_{13,3})
=\df16Z_{39}\label{d39}
\end{equation}
with $\theta_{13,3}:=2\arctan\sqrt{13/3}$. For $D=-84$,
we found that
\begin{equation}
60{\rm Cl}_2(\theta_{7,3})-36{\rm Cl}_2(2\theta_{7,3})
-4{\rm Cl}_2(3\theta_{7,3})+3{\rm Cl}_2(4\theta_{7,3})
+2{\rm Cl}_2(6\theta_{7,3})=\df16Z_{84}\label{d84}
\end{equation}
with $\theta_{7,3}:=2\arctan\sqrt{7/3}$.
These are our 7th and 8th unprovens, courtesy of Feynman.

\subsection{Two of Feynman's links are missing}

The rationale for the underlinings in~(\ref{iq}) should now be clear:
it was these 10 imaginary-quadratic fields that had been distinguished
by the Feynman orthoschemes of Table~22, which initiated our studies.
For 4 of these 10 fields, the census gave links;
for 2 of the remaining 6, the work in~\cite{9806174} gave links;
for 2 of the remaining 4, the alternating platonic link of Fig.~1
and the non-alternating daisy-chain link of Fig.~2
supplied answers; that left us with the $D=-39$ and $D=-84$ links
still missing.

It may be imagined that we spent much time looking
for Feynman's two missing links.
We find it significant that, in all our exploration,
SnapPea never detected a quadratic
field beyond those we have reported, with $-D\in\{3,4,7,8,11,15,20,24\}$.
One could, if one was minded, synthesize cusped manifolds
with gluing conditions that are satisfied in other quadratic fields,
and then assert the existence of links whose complements in $S^3$
are isometric to these manifolds.
Our aim was more concrete, and perhaps more old-fashioned.
Just as the basic ingredients of Table~22 would have been
immediately intelligible to their
originators\footnote{The geometry of Lobachevsky~(1792--1856) gives
a model for the Universe that accords with data.
Clausen~(1801--85) was an astronomer,
described by Gauss~(1777--1855) as a man of outstanding talents.}
Nikolai Ivanovich Lobachevsky and Thomas Clausen,
so would those of Figs.~1 and~2 have been
to\footnote{Maxwell~(1831--79) and Tait~(1831--1901)
were schoolfriends in Edinburgh. Maxwell later
formulated electromagnetic field theory and encouraged Tait's
search for a connection between knots and physics.}
James Clerk Maxwell and Peter Guthrie Tait.

It was, therefore, very satisfying to progress to $D=-24$,
at 48 crossings, and not to encounter any quadratic
SnapPea triangulation beyond those we had learnt
to expect from Feynman, Davydychev and Delbourgo.
By the same token, it was frustrating not to discover
two more Feynman links, with volumes rationally related to
$Z_{39}$ and $Z_{84}$.

We hope that others will be motivated to search. Table~19 indicates
the challenge. The self-dual platonic link of Fig.~1 entails 24 crossings,
10 components, and a volume
$Z_{20}={\rm vol}(24^{10}_{\rm tet})\approx50.447$;
the daisy-chain link of Fig.~2 entails 48 crossings and 24 components,
to reach $Z_{24}=V_{24}-V_{8}\approx62.186$.
The reader is left to imagine what might be entailed
by $Z_{39}\approx165.575$ and $Z_{84}\approx404.736$.

In any case, one now knows that
Feynman orthoschemes\footnote{The Feynman diagram evaluates to
the volume of orthoscheme~(\ref{Schl}) divided by a square root.}
at Mandelstam variables
$s=3t=13$ and $s=\frac32t=7$ are rationally related
to~(\ref{ZKq}) at $D=-39$ and $D=-84$, with coefficients
in~(\ref{d39},\ref{d84}) that are
as easy to measure, and as hard to prove, as those which are reified
in Figs.~1 and~2. It is hard to believe that the 2 missing links will
be less beautiful than the 8 which  we have already related to
Feynman orthoschemes.

\subsection{Dedekind zeta values from 10-crossing Feynman knots}

We found that 7 of the 84 knots with less than 10 crossings
have rational Dedekind-zeta invariants. The 6 distinct
values of~(\ref{ZK}) are
\begin{eqnarray}
Z_3&=&1\times{\rm vol}(4_1)\label{4_1}\\
Z_{23,3}&=&\df13\times{\rm vol}(5_2)
=\df{1}{10}\times{\rm vol}(9_{49})\label{5_2}\\
Z_{44,3}&=&\df13\times{\rm vol}(9_{48})\label{9_48}\\
Z_{59,3}&=&1\times{\rm vol}(7_{4})\label{7_4}\\
Z_{76,3}&=&1\times{\rm vol}(9_{35})\label{9_35}\\
Z_{448,4}&=&\df16\times{\rm vol}(8_{18})\label{8_18}
\end{eqnarray}
where the subscripts of $Z_{|D|,n}$ identify the
(negated) discriminant and degree of the number field,
and we omit the latter in the quadratic case.
Two further knots, $8_{21}$ and $9_{28}$, have invariant trace fields
in Table~18. From these subfields of joins, one may extract
\begin{eqnarray}
Z_7&=&4\times{\rm vol}(8_{21})-\df43\times{\rm vol}(8_{18})\label{8_21}\\
Z_{507,4}&=&\df25\times{\rm vol}(9_{28})-1\times{\rm vol}(4_1)\label{9_28}
\end{eqnarray}

We now report on two very special knots at 10 crossings.
Work begun by Dirk Kreimer~\cite{PLB,Hab,Sub},
and extended in collaborations with
Broadhurst, Delbourgo and Gracey~\cite{BKP,BDK,BGK,BK15},
has established a rich connection between multiple zeta
values~\cite{Eul,BBB,BBB1,BBB2} and positive knots, forged by multi-loop
Feynman diagrams in quantum field theory.
A positive knot is one with a minimal braid
word that entails exclusively positive powers of the generators of the
braid group~\cite{VJ}. There is an important
feature to note: no positive knot with less than 10 crossings
is hyperbolic. The 5 positive knots
with less than 10 crossings are the $(3,2)$, $(5,2)$, $(7,2)$,
$(4,3)$ and $(9,2)$
torus\footnote{Non-hyperbolic knots are torus or satellite knots,
with the latter beginning at 13 crossings.}
knots, with 3,~5,~7,~8 and~9 crossings, corresponding to
$\zeta(3)$, $\zeta(5)$, $\zeta(7)$, $\zeta(5,3)$
and $\zeta(9)$, where $\zeta(r,s):=\sum_{j>k>0}
j^{-r}k^{-s}$ is a multiple zeta value (MZV) of depth 2 and weight $r+s$,
with $\zeta(5,3)$ being the sole irreducible MZV below weight 10.

The first 2 hyperbolic Feynman knots are
$10_{139}:=\sigma_1^{}\sigma_2^{3}\sigma_1^{3}\sigma_2^{3}$
and $10_{152}:=\sigma_1^{2}\sigma_2^{2}\sigma_1^{3}\sigma_2^{3}$,
with volumes
\begin{eqnarray}
{\rm vol}(10_{139})&=&
4.8511707573327375670583270521153124788452830277699\ldots
\label{139}\\
{\rm vol}(10_{152})&=&
8.5360653472056086031441819205493259949649913969140\ldots
\label{152}
\end{eqnarray}
that are rather modest, compared with most of the
other 162 hyperbolic 10-crossing knots.

We do not know how many of the 164 hyperbolic 10-crossing knots
enjoy single-complex-place invariant trace fields, though we may estimate
the fraction from 3 previous results. We found rational relations
for $998/11031\approx9\%$ of the closed census manifolds, for
$312/4929\approx6\%$ of the cusped census manifolds, and
for $7/79\approx9\%$ of the hyperbolic knots with less than 10 crossings.
It thus seems unlikely that more than 10\% of 10-crossing knots
have rational Dedekind invariants. Had we selected a pair at random,
the odds on both having single-complex-place invariant trace fields
would be of order $100:1$ against. Yet the volumes~(\ref{139},\ref{152})
were not chosen at random; they come from the unique pair of positive
hyperbolic 10-crossing knots.

It is, therefore, both notable and gratifying that the Feynman knots
$10_{139}$ and $10_{152}$ both yield simple rational
Dedekind-zeta invariants, namely
$1$ and $1/2$, corresponding to the Dedekind zeta values
\begin{eqnarray}
\zeta_{K_{139}}(2)&=&1\times
\frac{(2\pi)^6{\rm vol}(10_{139})}{12\times688^{3/2}}\label{z139}\\
\zeta_{K_{152}}(2)&=&\frac12\times
\frac{(2\pi)^8{\rm vol}(10_{152})}{12\times8647^{3/2}}\label{z152}
\end{eqnarray}
for the single-complex-place quartic and quintic fields
\begin{eqnarray}
K_{139}&:=&x^4-2x-1\label{k139}\\
K_{152}&:=&x^5-3x^3-2x^2+2x+1\label{k152}
\end{eqnarray}
where $K_{139}$ is also the invariant trace field of the
link $8_2^2$, with the same volume as $10_{139}$.

There is an important distinction between the 2 hyperbolic
Feynman knots at 10 crossings and the sole hyperbolic Feynman
knot at 11 crossings, associated with the irreducible~\cite{Eul,BBB}
triple sum $\zeta(3,5,3)=\sum_{j>k>l>0}j^{-3}k^{-5}l^{-3}$.
At 10 crossings it has not yet been possible to identify
the generalized polylogarithms of weight 10 in the
7-loop\footnote{The loop number, $L$, is the number of
4-dimensional integrations over internal momenta; the crossing number
does not exceed $2L-3$. Numerical analysis of 28-dimensional integrals
is rather taxing.}
Feynman diagrams~\cite{BKP} that skein to gives these knots.
We know that each involves more than MZVs, since $\zeta(7,3)$
is accounted for by the $(5,3)$ torus knot, $10_{124}$

Plans are afoot to compute, to high precision, the numbers
associated by QFT to $10_{139}$ and $10_{152}$.
Discussion with Andrei Davydychev and Dirk Kreimer
suggests that it may be possible to extract values
from high-order $\varepsilon$-expansions of multi-loop
dressings of 2-loop skeletons in $4-2\varepsilon$ spacetime
dimensions, with dressings deliberately chosen to frustrate
reduction to MZVs.
If that project bears fruit, there will be scope for PSLQ
searches, beyond what is possible with current 10-digit
data from 7-loop diagrams.

It is believed that promising targets for such integer-relation
searches might be provided by volumes of polytopes in
hyperbolic spaces of odd dimensions substantially greater
than 3, and perhaps as large as 13,
which may present a computational challenge to geometers
as great as that confronting quantum field theorists at 7 loops.
However, it will do no harm to include in PSLQ searches
the easily computed weight-10 quintic Dedekind zeta
value~(\ref{z152}), from 3 dimensions, which would
correspond to the less likely hypothesis that the
associated geometry is simpler than appears from
the 7-loop physics.

We noted that at 9 crossings the 3 knots with rational Dedekind-zeta
invariants comprise the $3,3,3$ pretzel knot $9_{35}$, in~(\ref{9_35}),
and a pair of non-alternating knots, namely $9_{48}$ and $9_{49}$,
in~(\ref{5_2},\ref{9_48}). Accordingly, we sought for further
single-complex-place fields among the 10-crossing pretzels,
namely $10_{46}:=5,3,2$ and $10_{61}:=4,3,3$, and the
remaining non-alternating hyperbolic 10-crossing knots, namely
$\{10_n|165\ge n\ge125;\,n\neq139;\,n\neq152\}$. From these 41 knots
we obtained only 2 results:
\begin{eqnarray}
Z_{31,3}&=&
\df18\times{\rm vol}(10_{157})=
\df13\times{\rm vol}(7^2_1)\label{10_157}\\
Z_{29963,5}&=&4\times{\rm vol}(10_{153})\label{10_153}
\end{eqnarray}
providing a measure of how privileged is the Feynman~\cite{BKP,BK15} pair,
$10_{139}$ and $10_{152}$.
There was one join, already found in~(\ref{9_28}) at 9 crossings, with
\begin{equation}
Z_3+Z_{507,4}=\df12\times{\rm vol}(10_{155})=\df25\times{\rm vol}(9_{28})
\label{10_155}
\end{equation}
We note that the first example of degenerate volumes,
provided by
\begin{equation}
{\rm vol}(10_{132})={\rm vol}(9_{42})
\label{10_132}
\end{equation}
does not yield a rational invariant, since the quintic invariant trace fields
are generated by $x^5-x^4-2x^2+2x-1$, with 2 complex places.

We also remark on the alternating knot $10_{123}$. Like $9_{41}$,
it has a two-complex-place invariant trace field, generated by a root of
$x^4-x^3+x^2-x+1$. Thus its volume is provably related to Clausen values
at multiples of $\pi/5$. Interestingly, both knots have volumes that
are rationally related to instances of the orthoscheme~(\ref{Schl}), with
\begin{eqnarray}
S(\df25\pi,\df1{10}\pi,\df15\pi)&=&
\df{1}{10}\times{\rm vol}(9_{41})=
{\rm Cl}_2(\df25\pi)+\df13\,{\rm Cl}_2(\df45\pi)\label{9_41}\\
S(\df3{10}\pi,\df15\pi,\df1{10}\pi)&=&
\df{1}{10}\times{\rm vol}(10_{123})=
\df23\,{\rm Cl}_2(\df25\pi)+\df13\,{\rm Cl}_2(\df45\pi)\label{10_123}
\end{eqnarray}
The corresponding relations between Clausen values are straightforward
to derive, and hence quite unlike those with $-D\ge7$ in Table~22, which
involve the 8 dramatic switches between number fields recorded
in~(\ref{d7}--\ref{d20},\ref{d24},\ref{d39},\ref{d84}).

\subsection{Dedekind zeta values from 12-crossing Feynman knots}

At 12 crossings, corresponding to 8 loops in QFT,
Broadhurst and Kreimer~\cite{BK15} found that 7 of the 2,176 hyperbolic
knots are positive.
We found that 3 of these have rational Dedekind-zeta invariants, with
\begin{eqnarray}
Z_{23,3}&=&
\df13\times{\rm vol}(\sigma_1^{}\sigma_2^{3}\sigma_1^{}\sigma_2^{7})
\label{1317}\\
Z_{848,4}&=&
1\times{\rm vol}(\sigma_1^{}\sigma_2^{5}\sigma_1^{}\sigma_2^{5})
\label{1515}\\
Z_{2068,4}&=&
2\times{\rm vol}(\sigma_1^{3}\sigma_2^{3}\sigma_1^{3}\sigma_2^{3})
\label{3333}
\end{eqnarray}
The first two are those identified as Feynman knots in~\cite{BGK,BK15}.
Again, we find it uncanny that, with odds of at least $100:1$ against,
both Feynman knots proved to have single-complex-place invariant
trace fields. It is clear that the Dedekind single-complex-place
criterion and the Feynman positivity criterion are strongly related.
The origin of this association is, however, quite unclear to us.

Most notable is the result~(\ref{1317})
for the 12-crossing positive knot with braid word
$\sigma_1^{}\sigma_2^{3}\sigma_1^{}\sigma_2^{7}$, which is
associated~\cite{BGK}
with the irreducible MZV $\zeta(9,3)$ in QFT.
It has a rather small, and very special, volume:
\begin{equation}
{\rm vol}(\sigma_1^{}\sigma_2^{3}\sigma_1^{}\sigma_2^{7})
={\rm vol}(5_2)
=\df{3}{10}\times{\rm vol}(9_{49})
=\df37\times{\rm vol}(7^2_2)
=3Z_{23,3}
\approx2.828122\label{12A}
\end{equation}
which is precisely 3 times the volume of the closed Weeks manifold,
${\rm m}003(-3,1)$, conjectured to be the smallest of all hyperbolic
manifolds. The knot $\sigma_1^{}\sigma_2^{3}\sigma_1^{}\sigma_2^{7}$
is the first in the sequence of hyperbolic Feynman knots~\cite{BK15}
$F_{2n}:=\sigma_1^{}\sigma_2^{3}\sigma_1^{}\sigma_2^{2n-5}$,
with $2n\ge12$ crossings, associated
with $\zeta(2n-3,3)$ in Feynman
diagrams with $n+2\ge8$ loops.

We found that the equality
of the volumes of $F_{12}$ and $5_2$ generalizes to
\begin{equation}
{\rm vol}(\sigma_1^{}\sigma_2^{3}\sigma_1^{}\sigma_2^{2n-5})
={\rm vol}(\sigma_1^{}\sigma_2^{3}\sigma_1^{}\sigma_2^{11-2n})\label{vt}
\end{equation}
where the knot on the r.h.s.\ is formally obtained
by $n\to8-n$ and has no more than $2n-6$ crossings.
For $2n\ge12$ we also found that
\begin{equation}
{\rm vol}(F_{12})\le{\rm vol}(F_{2n})
\le{\rm vol}(9^2_{60})=\df12Z_7\approx5.333489\label{bound}
\end{equation}
where $Z_7$ is the pivot of~(\ref{d7}),
which is the first of the non-trivial Clausen relations
from 1-loop box diagrams.
At large $n$, the behaviour was measured to be
\begin{equation}
{\rm vol}(F_{2n})=\df12Z_7
-\frac{C}{(\frac14n-1)^2}+O(1/n^4)
\label{limits}
\end{equation}
with $C\approx0.8160$, found from 12-digit SnapPea results,
with up to 500 crossings.

The asymptote suggested that the manifolds complementary
to the series $F_{2n}$ of Feynman knots might be isometric to a series of
Dehn fillings of a manifold with volume $\frac12Z_7$.
The {\tt drill} command of Snap suggested the manifold ${\rm s}785$,
which we found to be isometric to the complement of the link $9^2_{60}$.
Performing the surgeries
$(-2,1)\ldots(-21,1)$ on its second cusp, we obtained 20 manifolds
and asked SnapPea to compare them with the manifolds complementary to
the Feynman knots $F_{2n}$, with crossing numbers from 12 to 50.
The result was isometry, in all 20 cases.
It was then possible to compute 64 good digits of
\begin{equation}
C={\tt
0.8160162119959694990691941006445603982758744790599736680521553757
\ldots}\label{C}
\end{equation}
from 20 high-precision Snap triangulations of ${\rm s}785(,)(4-n,1)$,
with $n=O(10^3)$. It would be interesting to obtain analytical results
for asymptotic changes~\cite{NZ} in volume, such as that given
by~(\ref{C}).

Now one sees the origin of the invariance~(\ref{vt})
of volumes, under the transformation $n\to8-n$. This merely flips a
sign of the Dehn surgery on the torus~\cite{Adams} curve $(4-n,1)$.
The fixed point, at $n=4$, is the non-hyperbolic
longitudinal surgery $(0,1)$, corresponding to the
$(4,3)$ torus knot, $F_8:=(\sigma_1^{}\sigma_2^{3})^2
\sim(\sigma_1\sigma_2)^4=8_{19}$,
which is the first 3-braid Feynman knot, found at 6 loops~\cite{BKP}
in QFT, where it signals the appearance of the first
irreducible~\cite{Eul,BBB} MZV, $\zeta(5,3)$, in the
counterterms of $\phi^4$-theory.
At 7 loops, with $n=5$, the surgery $(-1,1)$ is likewise
non-hyperbolic, corresponding to the $(5,3)$ torus knot
$F_{10}:=\sigma_1^{}\sigma_2^{3}\sigma_1^{}\sigma_2^{5}
\sim(\sigma_1^{}\sigma_2^{})^5=10_{124}$,
associated with $\zeta(7,3)$. Only at 12 crossings,
and hence 8 loops, does this series of Feynman knots
start to be hyperbolic. One thus expects to find a rather special volume
at 12 crossings, as is indeed seen in~(\ref{12A}).

\subsection{Dedekind zeta values from maximally symmetric knots}

Jim Hoste, Morwen Thistlethwaite and Jeff Weeks~(hereafter HTW)
have recently completed an impressive symmetry analysis~\cite{HTW}
of all 1,701,936 prime knots with up to 16 crossings. From the
University of Tennessee at Knoxville, we obtained
files\footnote{\tt
http://www.math.utk.edu/$\,\widetilde{}\,$morwen/knotscape.html}
that identify highly symmetric alternating and non-alternating
knots at crossing numbers from 11 to 16. Before analyzing the
most symmetric of these, we comment on the situation up to 10 crossings,
in territory charted by Dale Rolfsen~\cite{Rolfsen} and predecessors.

Table~23 gives the hyperbolic alternating and
non-alternating\footnote{Below 8 crossings, all hyperbolic knots are
alternating.} knots of maximal symmetry at crossing numbers from 4 to 10,
together with the symmetry groups, invariant trace fields, signatures
and discriminants of their complementary manifolds.
Where the field has a single complex
place (i.e.\ signature $[n-2,1]$ at degree $n$)
we give the rational Dedekind-zeta invariant, $a/b$.
Table~24 shows our remaining finds of Dedekind-zeta
invariants, for knots up to 10 crossings. The latter are
likely to be complete up to 9 crossings; at 10 crossings we analyzed
all non-alternating knots, but only about 20\% of the alternating knots.

Some comments are in order.
\begin{enumerate}
\item Up to 9 crossings, maximal symmetry is a good -- yet far from
infallible -- diagnostic of a single-complex-place field.
\item The maximally symmetric knots designated by Rolfsen~\cite{Rolfsen}
as $6_3$, $7_7$, $8_{21}$, $9_{40}$ and $10_{123}$ have invariant trace
fields with more than one complex place.
\item In~(\ref{8_21}) it is shown that the volume of
$8_{21}$ reduces to a pair of Dedekind zeta values.
\item In~(\ref{10_123}) it is shown that the volume of
$10_{123}$ reduces to that of a simple orthoscheme.
\item Only one Dedekind-zeta knot, namely $9_{49}$, escapes the
sieve of maximal symmetry up to 9 crossings. Moreover,~(\ref{5_2})
shows that it yields a Dedekind zeta value already encountered
at fewer crossings.
\item At 10 crossings, we found 3 non-alternating Dedekind-zeta knots
with less than maximal symmetry: the Feynman~\cite{BKP,BK15} pair,
$10_{139}$ and $10_{152}$, with the modest symmetries $D_2$ and $Z_2$,
and $10_{153}$, which SnapPea declared to be devoid of symmetry.
\end{enumerate}

In the light of the above, we were prepared for a dwindling yield
from maximal symmetry, above 10 crossings.
The harvest proved to be even more meagre than we anticipated.
Table~25 shows that {\sl none} of the 11 maximally
-- and visibly~\cite{HTW} -- symmetric alternating
knots from 11 to 16 crossings has a single-complex-place
invariant trace field. From the 10 maximally -- and
covertly~\cite{HTW} -- symmetric non-alternating knots,
we obtained 4 single-complex-place fields, yet only one of these
entailed a new Dedekind zeta value, namely
\begin{equation}
Z_{643,4}=\df15\times{\rm vol}({\rm n}14.13191)\label{643,4}
\end{equation}
while those in
\begin{eqnarray}
Z_{44,3}&=&\df14\,\times{\rm vol}({\rm n}12.642)\label{44,3}\\
Z_{448,4}&=&\df{1}{10}\times{\rm vol}({\rm n}15.112310)\label{448,4}\\
Z_{31,3}&=&\df{1}{15}\times{\rm vol}({\rm n}16.1007813)\label{31,3}
\end{eqnarray}
had already been obtained in~(\ref{9_48},\ref{8_18},\ref{10_157}).
The last of these duplicates merits further comment.

There are more~\cite{HTW}
than a million non-alternating hyperbolic 16-crossing knots. Amid this
plethora, HTW identified the 1,007,813th
(in their lexicographic ordering of Dowker codes) as uniquely
maximally symmetric.
It enjoys the 18-fold dihedral symmetry group $D_9$ of the nonagon,
exquisitely disguised in any 16-crossing projection.
If the reader has access to SnapPea, s/he should certainly not miss
the opportunity to click in a depiction of the non-alternating knot in
Fig.~7 of the highly readable article\footnote{{\tt
http://www.pitzer.edu/$\,\widetilde{}\,$jhoste}
also gives access to~\cite{HTW}.}
by HTW. Then the power of Jeff Weeks' topological engine~\cite{SnapPea}
becomes apparent, when its symmetry analyzer announces $D_9$.
Morwen Thistlethwaite's website\footnote{\tt
http://www.math.utk.edu/$\,\widetilde{}\,$morwen/d9.html}
renders this more visible, by resort to a 5-braid
presentation.

Intuition told us to expect a Dedekind zeta value from the HTW
knot with symmetry group $D_9$. In this respect, we were not disappointed:
the volume of the knot is $15\,Z_{31,3}$, giving a Dedekind-zeta
invariant $a/b=1/15$ that is smaller than any we had previously encountered.
We allayed the disappointment,
at having already encountered $Z_{31,3}$, by the following
{\sl ex post facto} considerations.
The $D_9$ knot is so special that it merits a simple
invariant trace field. The quadratic field ${\bf Q}(\sqrt{-3})$
was taken by the figure-8 knot, in~(\ref{4_1}).
The cubic field with $D=-23$ had already been engaged
by Feynman, at 12 crossings, in~(\ref{12A}).
The next cubic discriminant, $D=-31$, provides an
eminently suitable resting place for the Hoste--Thistlethwaite--Weeks
$D_9$ knot. Colleagues engaged on commensurability~\cite{CGHN}
analyses may now investigate the wondrously long chain
\begin{eqnarray}
Z_{31,3}
&=&\df{1}{15}\times{\rm vol}({\rm n}16.1007813)\label{1/15}\\
&=&\df18\,\times{\rm vol}(10_{157})\label{1/8}\\
&=&\df14\,\times{\rm vol}({\rm v}3183)\label{1/4}\\
&=&\df13\,\times{\rm vol}(7^2_1)\label{1/3}\\
&=&\df12\,\times{\rm vol}({\rm m}034)\label{1/2}\\
&=&1\,\,\times{\rm vol}({\rm m}007(+4,1))\label{1/1}\\
&=&\df23\,\times{\rm vol}({\rm m}149(+1,2))\label{2/3}\\
&=&\df27\,\times{\rm vol}({\rm v}1963)\label{2/7}
\end{eqnarray}
of distinct rational Dedekind-zeta invariants
from a common invariant trace field. The last 6 invariants cover
43 census manifolds. The volumes of the
$D_9$ knot and $10_{157}$ are $30\,D(z)$ and $16\,D(z)$,
where $z^3=1-z$, with $\Im z>0$, and $D(z)=\frac12Z_{31,3}$
is the volume of any one of the 4 ideal
tetrahedra that triangulate ${\rm m}034$. Here, as in the
Feynman case~(\ref{12A}), the Dedekind zeta value collapses
to a rational multiple of a single Bloch--Wigner dilogarithm.

\subsection{Dedekind zeta value from an 18-crossing Feynman knot}

It was observed that the positivity criterion of~\cite{BKP,BGK,BK15}
was more fertile than maximal symmetry, at 10 and 12 crossings.
We expect further sequences of positive knots
whose first hyperbolic instances yield Dedekind zeta values
at high crossing numbers. From QFT
we inferred a source of such a sequence, namely the 3-parameter
family of even-crossing positive 3-braids~\cite{BK15}
\begin{equation}
R_{k,m,n}:=
\sigma_1^{}\sigma_2^{2k}\sigma_1^{}\sigma_2^{2m}\sigma_1^{}\sigma_2^{2n+1}
\label{BK15R}
\end{equation}
We knew from~\cite{BK15} that $R_{2,2,1}$ is the 14-crossing torus knot
$(7,3)$. Dirk Kreimer helped us show that $R_{3,2,1}$ is the 16-crossing
torus knot $(8,3)$. But $R_{4,2,1}$ cannot be a torus knot,
since 9 and 3 are not coprime. The 18-crossing hyperbolic
volume is intriguingly small:
\begin{equation}
{\rm vol}(R_{4,2,1})={\tt
3.47424776131274229602900855361193191879781770805621
\ldots}\label{vR412}
\end{equation}
Table~13b immediately identified the invariant trace field
as a single-complex-place sextic, with $D=-753079$. Table~6 then
gave $x^6 - x^5 - 3x^4 - x^3 + 2x^2 + 2x - 1$ as the
generating polynomial. SnapPea confirmed isometry of the
 complement of $R_{4,2,1}$ with manifold ${\rm m}082$.
The corresponding rational Dedekind-zeta invariant, $a/b=26$, in
\begin{equation}
Z_{753079,6}=26\times{\rm vol}(R_{4,2,1})
\label{46}
\end{equation}
is larger than we had found for any graphically generated knot,
and is 390 times that for the $D_9$ knot, in~(\ref{1/15}). For the 5th time
of asking, a positive Feynman knot gives a Dedekind zeta value.
The odds on this being accidental are at least $10^5:1$ against.

Lest such a connection between knots, numbers and
Feynman diagrams be thought exceptional, we recall that at 8
crossings QFT demanded~\cite{BKP} a positive 3-braid
knot and an irreducible depth-2 MZV, which were duly forthcoming,
in the shape of $8_{19}$ and $\zeta(5,3)$. At 11 crossings, the demands
were similarly imperious: a positive 4-braid knot and an irreducible depth-3
MZV, satisfied by the uniqueness~\cite{BKP} of
$\sigma_1^{}\sigma_2^{3}\sigma_3^{2}\sigma_1^{2}\sigma_2^{2}\sigma_3^{}$
and the irreducibility~\cite{Eul} of $\zeta(3,5,3)$.
At 12 crossings, QFT seemed, at first, to require
something perverse: an arbitrary choice between an irreducible
depth-4 MZV and an irreducible depth-2
alternating~\cite{Eul} Euler sum. Mathematics accommodated, via the
remarkable discovery~\cite{BGK,BK15,Eul} that one is reducible to the other.
Compared with these past findings,
a new Dedekind zeta value, at 18 crossings,
is small fry to the maw of natural philosophy.

It took 50 years to discover that the renormalization of QFT
is governed by a Hopf algebra~\cite{HA}, richer than that
of noncommutative geometry~\cite{CK}, and readily automated~\cite{ESI}.
This offers the prospect of elucidating the existence
of analytically non-trivial 4-term relations~\cite{4TR,4TT} in
realistic (i.e.\ 4-dimensional) QFT.
Hopefully, our latest addition of a Dedekind-zeta connection,
to the melting-pot of knot/number/field theory~\cite{DKbook},
may also be illuminated by the joint efforts of physicists and
mathematicians who are responsive to empirical data.

\section{Conclusions and prospects}

Perhaps more than in any other piece of research which
either author has undertaken,
this work has been vitally enabled by the internet.
It provided us with the opportunity to blend number theory,
topology, geometry, analysis, physics and computer science,
in a global empirical mixing bowl, thanks to:
\begin{itemize}
\item ready access to significant data at
Bordeaux, Claremont CA, Florham Park NJ, Knoxville TN, Melbourne,
Minneapolis MN,
as detailed in footnotes;
\item the ability to run high-level packages, such as Maple, Pari
and Reduce, on whatever machine best served our purposes, irrespective
of the contingencies of our personal geographic co-ordinates;
\item wonderful specialized resources, downloadable as per our footnotes,
namely: David Bailey's superb PSLQ and FFT routines,
Oliver Goodman's port of high-precision Snap to DigitalUnix,
and Al Manoharan's attractive adaptation to Windows95 of Jeff Weeks'
amazing SnapPea program,
all supported by generous email advice;
\item access to powerful computers in England,
Newfoundland\footnote{{\tt
http://www.cecm.sfu.ca/$\,
\widetilde{}\,$jborwein/PUP\_report\_March15/report.html}} and Vancouver,
yielding high-precision results
such as those in~(\ref{50dp}) and~(\ref{C}),
and the 1800-digit hyperbolic
volumes of Tables~20 and~21, achieved by dedicated
multiple-precision~\cite{TRANSMP} code,
after exploratory work enabled by the above.
\end{itemize}

The facility with which we were able to plug into all these valuable
resources advertises how rich the opportunities for empirical mathematics
are becoming. The tools came together to offer more patterns
than we had dared to hope for.
Enterprises such as we have limned promise to be more and more a
part of mathematical and physical research in the next few decades.

That said, many of the results which we have exhibited remain
tantalizingly far from proof, let alone understanding. Here we repeat
4 of many remaining puzzles:
\begin{itemize}
\item  How might one begin to derive relations such as~(\ref{modern}),
between dilogarithms with arguments in radically different number fields?
\item Why do 1-loop Feynman diagrams, at very specific values of the
Mandelstam variables, generate even more relations than we were
able to reify by quadratic links?
\item Why do Feynman diagrams, with 7, 8 and 11 loops, lead to 5
Dedekind-zeta knots, at 10, 12 and 18 crossings, with odds of a
chance association being at least $10^5:1$ against?
\item Can the mere dilogarithms of 3-dimensional hyperbolic geometric
tell us anything about the unidentified
10th-order polylogarithms of 7-loop quantum field theory?
\end{itemize}

When faced by such visible expansion of one's lack of wisdom,
it is probably best to concentrate on that which is easiest to state.
Hence we conclude with a rewriting of the simplest
unproven relation~(\ref{modern}) in terms of the
Dirichlet series~(\ref{chi}). From light-by-light scattering at $s=t=7$,
and -- just as mysteriously -- from the hyperbolic volume
of the link $6^3_1$, we infer that
\begin{eqnarray}
&&\sum_{n>0}\left\{
 3\left(-\frac34+\frac{\sqrt{-7}}{4}\right)^n
-3\left(-\frac34+\frac{\sqrt{-7}}{4}\right)^{2n}
 +\left(-\frac34+\frac{\sqrt{-7}}{4}\right)^{3n}\right\}\frac{1}{n^2}
\nonumber\\
&&\qquad\qquad{}=13\,\zeta(2)-6\,\pi \,\arctan\sqrt {7}
+\frac{7\sqrt{-7}}{4}
\sum_{n>0}\left(\frac{n}{7}\right)\frac{1}{n^2}
\label{finis}
\end{eqnarray}
The real part of this empirical and indubitable equality is
easily proven; in its imaginary part, with the Jacobi symbol
$\left(\frac{n}{7}\right)$, resides a flinty kernel which
is -- to us at least -- intractable.
A proof of (\ref{finis}) and the other quadratic identities
would be most welcome.

\subsection*{Acknowledgments}

We thank David Bailey, Oliver Goodman, and Al Manoharan,
for adapting PSLQ, Snap, and SnapPea to our chosen platforms;
without their personal help our work would not have been completed.
Equally vital was the contribution of Andrei Davydychev and
Bob Delbourgo, without whose ideas it would not have begun.
Advice and encouragement from Petr Lisonek, during DJB's
visit to CECM, were much appreciated.
JMB thanks Al Manoharan for discussions at MSRI, Berkeley.
DJB thanks Paul Clark, for explaining the difference between
geometry and topology, David Bailey and Helaman Ferguson, for discussions
at NERSC, Berkeley, which wedded computational architecture
to hyperbolic sculpture, and Alain Connes and Ivan Todorov,
for the {\sl Number Theory in Physics} workshop
at the Erwin Schr\"odinger Institute in Vienna,
where discussions with Paula Cohen, Dirk Kreimer and Don Zagier
encouraged completion of Tables~8--12.
We thank Dirk Kreimer for close readings of preliminary drafts,
and most of all for shaping our context.

\newpage

\begin{center}{\bf Table 1:}\quad Dedekind zeta values
hereby related to volumes of closed manifolds\end{center}
$$\begin{array}{|l|rrrrrrrrrrr|r|}\hline
\mbox{degree}&2&3&4&5&6&7&8&9&10&11&12&\le12\\\hline
\mbox{number}&3&11&32&38&25&19&5&3&2&1&1&140\\\hline\end{array}$$

\bet{2}{Quadratic fields}
\tabw{x^2-x+1}{3}{2}{1}{{\rm m}007(+3,1)}
\tabw{x^2+1}{4}{2}{1}{{\rm m}009(+5,1)}
\tabw{x^2-x+2}{7}{4}{1}{{\rm m}036(-4,3)}
\ent

\bet{3}{Cubic fields}
\tabw{x^3-x^2+1}{23}{1}{1}{{\rm m}003(-3,1)}
\tabw{x^3+x-1}{31}{1}{1}{{\rm m}007(+4,1)}
\tabw{x^3-x^2+x+1}{44}{2}{1}{{\rm m}006(+3,1)}
\tabw{x^3+2x-1}{59}{4}{1}{{\rm m}004(+6,1)}
\tabw{x^3-2x-2}{76}{2}{1}{{\rm s}784(+1,2)}
\tabw{x^3-x^2+x-2}{83}{4}{1}{{\rm m}034(-3,2)}
\tabw{x^3-x^2+2x+1}{87}{2}{1}{{\rm s}784(-1,2)}
\tabw{x^3-x-2}{104}{4}{1}{{\rm s}297(+1,3)}
\tabw{x^3-x^2+3x-2}{107}{4}{1}{{\rm m}168(-3,2)}
\tabw{x^3-x^2-2}{116}{4}{1}{{\rm s}881(-1,3)}
\tabw{x^3-x^2+x+2}{139}{4}{1}{{\rm v}3106(+3,1)}
\ent

\newpage

\bet{4}{Quartic fields}
\tabw{x^4-x^3+2x-1}{275}{2}{5}{{\rm m}016(-4,3)}
\tabw{x^4-x-1}{283}{1}{1}{{\rm m}003(-2,3)}
\tabw{x^4-x^3+x^2+x-1}{331}{1}{1}{{\rm m}003(-4,3)}
\tabw{x^4-x^2-1}{400}{2}{5}{{\rm m}400(+4,1)}
\tabw{x^4-2x^3+x^2-2x+1}{448}{1}{1}{{\rm m}010(+3,2)}
\tabw{x^4-x^3-x^2+3x-1}{491}{1}{1}{{\rm m}029(-3,2)}
\tabw{x^4-x^3-x^2-x+1}{507}{1}{1}{{\rm m}160(-3,2)}
\tabw{x^4-x^3+x^2-x-1}{563}{1}{1}{{\rm m}130(+1,4)}
\tabw{x^4-x^3-2x+1}{643}{1}{1}{{\rm m}247(-1,4)}
\tabw{x^4-2x-1}{688}{2}{1}{{\rm m}019(+3,4)}
\tabw{x^4-x^3+2x^2-1}{731}{1}{1}{{\rm s}649(-3,4)}
\tabw{x^4-2x^3+x^2-x-1}{751}{2}{1}{{\rm m}081(-4,1)}
\tabw{x^4-x^2-2x+1}{848}{2}{1}{{\rm m}207(-1,3)}
\tabw{x^4-2x^3+3x^2-1}{976}{2}{1}{{\rm m}286(-5,1)}
\tabw{x^4-x^3+x^2-3x+1}{1099}{2}{1}{{\rm s}663(+1,2)}
\tabw{x^4-x^3-2x-1}{1107}{2}{1}{{\rm s}928(+4,1)}
\tabw{x^4-x^3-2x^2-x+1}{1156}{4}{1}{{\rm m}082(+1,3)}
\tabw{x^4-x^3+2x^2+x-1}{1192}{4}{1}{{\rm m}148(-3,2)}
\tabw{x^4-x^2-3x-1}{1255}{2}{1}{{\rm v}3492(+4,3)}
\tabw{x^4+2x^2-x-1}{1371}{2}{1}{{\rm v}3489(+2,3)}
\tabw{x^4-x^3+x-2}{1399}{4}{1}{{\rm m}293(+2,3)}
\tabw{x^4-x^3-3x^2+2}{1588}{8}{1}{{\rm m}038(+4,1)}
\tabw{x^4-x^3+3x-1}{1732}{4}{1}{{\rm v}3187(-4,1)}
\tabw{x^4-x^3-x^2-2x+1}{1791}{4}{1}{{\rm s}961(+2,3)}
\tabw{x^4-x^3-2x^2-3x+1}{1879}{4}{1}{{\rm v}2914(+2,3)}
\tabw{x^4-2x^3+x^2-3x+1}{1927}{4}{1}{{\rm v}3452(-5,1)}
\tabw{x^4-4x^2-2x+2}{1968}{6}{1}{{\rm s}594(+1,3)}
\tabw{x^4-x^3-2x^2+3x+1}{2068}{8}{1}{{\rm m}389(+3,1)}
\tabw{x^4-x^2-3x-2}{2151}{12}{1}{{\rm m}015(+8,1)}
\tabw{x^4-x^3-2x^2+3x+2}{2319}{6}{1}{{\rm v}2944(-5,2)}
\tabw{x^4-2x^3-x^2+2x-2}{3312}{12}{1}{{\rm v}3209(+2,3)}\hline
\tabw{x^4-5x^2-4}{6724}{64}{1}{{\rm s}479(-5,1)}
\ent

\newpage

\bet{5}{Quintic fields}
\tabw{x^5-x^3-2x^2+1}{4511}{1}{1}{{\rm m}003(-5,3)}
\tabw{x^5-x^4-x^3+2x^2-x-1}{4903}{1}{1}{{\rm m}015(+5,1)}
\tabw{x^5-x^4-x^3+3x^2-1}{5519}{1}{1}{{\rm m}016(+3,2)}
\tabw{x^5-2x^4+x^3+2x^2-2x-1}{5783}{1}{1}{{\rm m}016(+2,3)}
\tabw{x^5-x^3-x^2-x+1}{7031}{1}{1}{{\rm m}160(-4,1)}
\tabw{x^5-2x^4+x^3-2x+1}{7463}{1}{1}{{\rm m}178(+4,3)}
\tabw{x^5-3x^3-2x^2+2x+1}{8647}{2}{1}{{\rm m}016(+4,1)}
\tabw{x^5-x^4-x^3+x^2-2x+1}{9439}{1}{1}{{\rm v}2759(-3,1)}
\tabw{x^5-2x^4+x^2-2x-1}{9759}{3}{1}{{\rm m}007(-3,2)}
\tabw{x^5-x^4-3x^3+3x-1}{10407}{3}{1}{{\rm m}023(-4,1)}
\tabw{x^5-x^4-2x+1}{11243}{2}{1}{{\rm s}090(+5,1)}
\tabw{x^5-x^4-2x^3+3x^2-x-1}{11551}{2}{1}{{\rm m}223(-1,3)}
\tabw{x^5+x^3-x^2-3x-1}{12447}{2}{1}{{\rm s}657(-1,2)}
\tabw{x^5-2x^2-x+1}{13219}{2}{1}{{\rm s}645(+1,3)}
\tabw{x^5-x^4-x^3-x^2-3x+1}{13523}{2}{1}{{\rm v}2530(+1,3)}
\tabw{x^5-x^4-2x^3+x+2}{13883}{2}{1}{{\rm v}3310(+5,1)}
\tabw{x^5-2x^4+2x^3-x^2-2x+1}{14103}{2}{1}{{\rm s}784(+5,2)}
\tabw{x^5-x^4-3x^3+3x^2-1}{14631}{2}{1}{{\rm s}958(+3,2)}
\tabw{x^5-x^4-2x^3-x^2+3x+1}{14911}{2}{1}{{\rm v}2704(-5,1)}
\tabw{x^5-x^4-2x^3-x^2+2x+2}{17348}{8}{1}{{\rm m}006(-5,1)}
\tabw{x^5-2x^4+4x^2-x-1}{22331}{4}{1}{{\rm s}884(+2,3)}
\tabw{x^5-2x^3-x^2-x+1}{22424}{4}{1}{{\rm v}3246(-2,3)}
\tabw{x^5-x^4+x^2-3x+1}{23103}{6}{1}{{\rm m}223(+5,1)}
\tabw{x^5-x^4+2x^2-2x-1}{23339}{4}{1}{{\rm v}3199(+3,1)}
\tabw{x^5-2x^3-3x^2+x+1}{29444}{8}{1}{{\rm s}478(-1,3)}
\tabw{x^5-2x^4-2x^3+4x^2-x+1}{29963}{16}{1}{{\rm m}007(-5,1)}
\tabw{x^5-x^4+3x^2-6x+2}{31684}{8}{1}{{\rm v}2381(+3,1)}
\tabw{x^5-2x^4+2x^3+x^2-3x-1}{34436}{8}{1}{{\rm v}2794(-2,3)}
\tabw{x^5-2x^4+2x^3-3x^2-x+4}{34779}{9}{1}{{\rm v}3214(+3,1)}
\tabw{x^5-2x^4-x^3+2x^2-x+3}{38083}{14}{1}{{\rm s}437(+1,3)}
\tabw{x^5-2x^4-x^3+4x^2-2x-2}{58064}{40}{1}{{\rm m}141(+2,3)}\hline
\tabw{x^5-x^4+3x^2-8x+4}{60803}{28}{1}{{\rm m}148(+5,1)}
\tabw{x^5-2x^4-2x^2+4}{70736}{32}{1}{{\rm v}2787(-1,3)}
\tabw{x^5-5x^3-2x^2+3x+2}{79952}{36}{1}{{\rm s}594(+3,2)}
\tabw{x^5-2x^4-2x^3+4x^2-x-2}{107264}{52}{1}{{\rm v}3454(-5,1)}
\tabw{x^5-2x^4-3x^3+x^2+5x+2}{112919}{88}{1}{{\rm m}304(+1,3)}
\tabw{x^5-x^4-x^3-6x^2-7x-2}{141791}{104}{1}{{\rm s}707(+5,1)}
\tabw{x^5-6x^3-5x^2-2x-4}{239639}{184}{1}{{\rm s}918(+3,2)}
\ent

\newpage

\bet{6}{Sextic fields}
\tabw{x^6-x^5-2x^4+3x^3-x^2-2x+1}{92779}{1}{1}{{\rm m}222(-6,1)}
\tabw{x^6-2x^4-2x^3+3x+1}{94363}{1}{1}{{\rm m}345(+1,2)}
\tabw{x^6-x^5-x^4+2x^3-2x^2-x+1}{104483}{1}{1}{{\rm s}648(+1,2)}
\tabw{x^6-2x^5+x^4-2x^3-x^2+3x-1}{161939}{2}{1}{{\rm s}682(+3,1)}
\tabw{x^6-x^5-2x^4-2x^3+x^2+3x+1}{215811}{6}{1}{{\rm m}015(-3,2)}
\tabw{x^6-x^5-4x^4+4x^3+4x^2-2x-1}{238507}{7}{1}{{\rm m}034(+4,1)}
\tabw{x^6-x^5-2x^4-3x^3+3x^2+3x-2}{365263}{26}{1}{{\rm m}004(+3,2)}
\tabw{x^6-2x^5-2x^4+6x^3-2x^2-5x+3}{417467}{13}{2}{{\rm v}3375(-5,2)}
\tabw{x^6-2x^5-x^4+5x^3-3x^2-3x+2}{463471}{16}{1}{{\rm m}249(+4,1)}
\tabw{x^6-2x^5-2x^4+5x^3-3x^2+3x-1}{561863}{10}{1}{{\rm v}3239(+3,2)}
\tabw{x^6-2x^5+6x^3-3x^2-2x+1}{629952}{18}{1}{{\rm s}386(+5,2)}\hline
\tabw{x^6-x^5-3x^3-3x^2+5x-1}{661831}{16}{1}{{\rm v}3361(+1,3)}
\tabw{x^6-x^5-x^4+5x^3+x^2-3x-1}{688927}{14}{1}{{\rm v}3243(-3,1)}
\tabw{x^6-4x^4-2x^3+3x^2+5x+1}{709783}{20}{1}{{\rm s}952(-4,1)}
\tabw{x^6-x^5-3x^4-x^3+2x^2+2x-1}{753079}{26}{1}{{\rm m}337(-3,1)}
\tabw{x^6-2x^5-2x^4+5x^3-2x^2-3x+2}{899447}{44}{1}{{\rm m}189(+3,2)}
\tabw{x^6-2x^5-2x^4+7x^3-x^2-5x+1}{1014119}{38}{1}{{\rm m}286(-6,1)}
\tabw{x^6-x^5+5x^3-4x^2-4x+2}{1107052}{50}{1}{{\rm v}1315(-4,1)}
\tabw{x^6-2x^5-2x^4+4x^3-2x^2-2x+1}{1290496}{80}{1}{{\rm m}285(-4,1)}
\tabw{x^6-2x^5+x^4-3x^3-x^2+5x-2}{1494223}{56}{1}{{\rm v}3036(+3,2)}
\tabw{x^6-x^5-3x^4+5x^3+x^2-4x-1}{1825672}{100}{1}{{\rm m}358(-5,3)}
\tabw{x^6-9x^3-10x^2-x+1}{2803244}{236}{1}{{\rm m}192(-5,2)}
\tabw{x^6-x^5-4x^4+8x^3+9x^2-10x-7}{4241707}{296}{1}{{\rm v}3428(-4,1)}
\tabw{x^6+x^4-7x^3-2x^2+7x+2}{5873596}{688}{1}{{\rm v}1858(+6,1)}
\tabw{x^6-x^5-3x^4+8x^3-3x^2-7x+1}{7792864}{976}{1}{{\rm v}2789(-2,3)}
\ent

\newpage

\bet{7}{Septic fields}
\tabw{x^7-x^6-x^5+4x^4-2x^3-4x^2+x+1}
{3685907}{14}{1}{{\rm m}006(-3,2)}
\tabw{x^7-x^6-2x^5+5x^4-6x^2+x+1}
{3998639}{17}{1}{{\rm m}004(+5,2)}
\tabw{x^7-3x^5-3x^4+4x^3+5x^2-2x-1}
{4297259}{10}{1}{{\rm m}221(-1,2)}
\tabw{x^7-2x^5-3x^4-3x^3+3x^2+4x+1}
{4795631}{13}{1}{{\rm m}032(+5,2)}
\tabw{x^7-2x^6-x^5+7x^4-5x^3-5x^2+5x-1}
{6515927}{11}{1}{{\rm s}900(+3,2)}
\tabw{x^7-2x^6-3x^5+3x^4+5x^3-x^2-3x+1}
{7215127}{46}{1}{{\rm m}004(+1,2)}
\tabw{x^7-2x^6+4x^4-5x^3-2x^2+4x+1}
{7557047}{14}{1}{{\rm v}2221(-1,3)}
\tabw{x^7-x^6-5x^5+6x^4+6x^3-5x^2-2x+1}
{7729991}{32}{1}{{\rm m}038(+1,2)}
\tabw{x^7-2x^6-4x^5+6x^4+6x^3-4x^2-3x+1}
{9429911}{20}{1}{{\rm s}838(-2,3)}\hline
\tabw{x^7-3x^6-x^5+8x^4-4x^3-3x^2+5x-2}
{12558899}{68}{1}{{\rm m}070(-3,2)}
\tabw{x^7-x^6-7x^5+6x^4+6x^3-11x^2+3x+2}
{32775179}{316}{1}{{\rm m}026(-5,2)}
\tabw{x^7-2x^6-4x^5+8x^4+2x^3-7x^2+5x+1}
{43210364}{242}{1}{{\rm v}3305(-1,2)}
\tabw{x^7-2x^6-2x^5+8x^4-2x^3-10x^2+x+3}
{46692071}{250}{1}{{\rm v}2825(-4,1)}
\tabw{x^7-3x^6-x^5+12x^4-9x^3-10x^2+8x+1}
{50052727}{296}{1}{{\rm v}2200(-3,2)}
\tabw{x^7-3x^6-4x^5+15x^4-11x^2+x+2}
{58360112}{788}{1}{{\rm m}034(+5,2)}
\tabw{x^7-2x^6-3x^5+x^4+8x^3+5x^2-6x-3}
{66467451}{480}{1}{{\rm s}523(-6,1)}
\tabw{x^7-3x^6-2x^5+11x^4-4x^3-6x^2+4x-2}
{75117248}{616}{1}{{\rm v}1788(+3,2)}
\tabw{x^7-2x^6-6x^5+4x^4+10x^3-x^2-4x+1}
{81589747}{828}{1}{{\rm m}213(-5,2)}
\tabw{x^7-2x^6-x^5-2x^4+2x^3+11x^2+2x-2}
{97569124}{736}{1}{{\rm v}3214(+2,3)}
\ent

\newpage

\bet{8}{Octadic fields}
\tabb{x^8-6x^6-5x^5+7x^4}\tabw{+10x^3-x^2-4x-1}
{202734487}{100}{1}{{\rm m}038(+3,2)}
\tabb{x^8-x^7-3x^6+7x^5-2x^4}\tabw{-8x^3+4x^2+2x-1}
{948381887}{496}{1}{{\rm s}900(-2,3)}
\tabb{x^8-2x^7-3x^6+11x^5-4x^4}\tabw{-13x^3+8x^2+5x-2}
{2095218667}{2240}{1}{{\rm s}901(-3,2)}
\tabb{x^8-3x^7-2x^6+16x^5-11x^4}\tabw{-21x^3+16x^2+12x-1}
{2401259831}{1978}{1}{{\rm v}3184(+4,1)}
\tabb{x^8-4x^7-4x^6+14x^5+23x^4}\tabw{-13x^3-32x^2-6x+4}
{81051965432}{636704}{1}{{\rm v}3109(-2,3)}
\ent

\bet{9}{Nonadic fields}
\tabb{x^9-3x^8-2x^7+11x^6-5x^5}\tabw{-10x^4+11x^3+x^2-4x+1}
{8843652791}{571}{1}{{\rm m}115(-5,2)}
\tabb{x^9-3x^8-4x^7+16x^6+x^5}\tabw{-22x^4+6x^3+4x^2-4x+4}
{48502810352}{5230}{1}{{\rm v}3157(+5,1)}
\tabb{x^9-3x^8-3x^7+13x^6}\tabw{-13x^4+2x^3-2x^2+3x+1}
{99961920379}{15436}{1}{{\rm s}649(-5,3)}
\ent

\bet{10}{Decadic fields}
\tabb{x^{10}-4x^8-5x^7+5x^6+19x^5}
\tabw{-2x^4-21x^3+x^2+6x-1}
{271488204251}{3669}{1}{{\rm m}006(-5,2)}
\tabb{x^{10}-4x^9-x^8+20x^7-15x^6-23x^5}
\tabw{+29x^4-4x^3-7x^2+6x-1}
{7748687650003}{232080}{1}{{\rm s}900(+2,3)}
\ent

\bet{11}{Endecadic field}
\tabb{x^{11}-3x^{10}-5x^9+20x^8+3x^7-42x^6}
\tabw{+14x^5+28x^4-17x^3-x^2+4x-1}
{21990497831723}{68838}{1}{{\rm m}007(-5,2)}
\ent

\bet{12}{Duodecadic field}
\tabb{x^{12}-3x^{11}-8x^{10}+17x^9}
\tabb{+27x^8-19x^7-50x^6-24x^5}
\tabw{+44x^4+37x^3-5x^2-8x-1}
{12476239474594496}{9408656}{1}{{\rm v}2824(+4,1)}
\ent

\newpage

\begin{center}
{\bf Table~13a:}\quad Rational relations of Dedekind zeta values to volumes
\end{center}
$$\begin{array}{|rr|rr|r|r|}\hline-D&n&a&b&
{\cal M}\qquad{}&{\rm vol}({\cal M})\qquad\qquad\qquad{}\\\hline
                  23&  3&         1&   1& {\rm m}003(-3,1) &
                   0.942707362776927720921299603\\
                 283&  4&         1&   1& {\rm m}003(-2,3) &
                   0.981368828892232088091452189\\
                   3&  2&         2&   1& {\rm m}007(+3,1) &
                   1.014941606409653625021202554\\
                 331&  4&         1&   1& {\rm m}003(-4,3) &
                   1.263709238658043655884716346\\
                  59&  3&         4&   1& {\rm m}004(+6,1) &
                   1.284485300468354442460337084\\
             7215127&  7&        46&   1& {\rm m}004(+1,2) &
                   1.398508884150806640509594326\\
                  23&  3&         2&   3& {\rm m}009(+4,1) &
                   1.414061044165391581381949404\\
              365263&  6&        26&   1& {\rm m}004(+3,2) &
                   1.440699006727364875282370223\\
             3998639&  7&        17&   1& {\rm m}004(+5,2) &
                   1.529477329430026262824928629\\
                4511&  5&         1&   1& {\rm m}003(-5,3) &
                   1.543568911471855074328472943\\
                  31&  3&         1&   1& {\rm m}007(+4,1) &
                   1.583166660624812836166028851\\
                  44&  3&         2&   1& {\rm m}006(+3,1) &
                   1.588646639300162988176913812\\
             3685907&  7&        14&   1& {\rm m}006(-3,2) &
                   1.649609715808664120798395881\\
                4903&  5&         1&   1& {\rm m}015(+5,1) &
                   1.757126029188451362874746593\\
                9759&  5&         3&   1& {\rm m}007(-3,2) &
                   1.824344322202911961274957217\\
                   4&  2&         2&   1& {\rm m}009(+5,1) &
                   1.831931188354438030109207029\\
               29963&  5&        16&   1& {\rm m}007(-5,1) &
                   1.843585972326677938720454754\\
                  23&  3&         1&   2& {\rm m}016(-3,2) &
                   1.885414725553855441842599206\\
               17348&  5&         8&   1& {\rm m}006(-5,1) &
                   1.941503084027467793730320127\\
                 283&  4&         1&   2& {\rm m}006(+2,3) &
                   1.962737657784464176182904379\\
               10407&  5&         3&   1& {\rm m}023(-4,1) &
                   2.014336583776842504278826477\\
        271488204251& 10&      3669&   1& {\rm m}006(-5,2) &
                   2.028853091474922845797067756\\
                   3&  2&         1&   1& {\rm m}036(-3,2) &
                   2.029883212819307250042405108\\
                 448&  4&         1&   1& {\rm m}010(+3,2) &
                   2.058484368193033362456050739\\
      21990497831723& 11&     68838&   1& {\rm m}007(-5,2) &
                   2.065670838488380741576307932\\
                5519&  5&         1&   1& {\rm m}016(+3,2) &
                   2.114567693110222238090213031\\
                8647&  5&         2&   1& {\rm m}016(+4,1) &
                   2.134016336801402150786045480\\
              238507&  6&         7&   1& {\rm m}034(+4,1) &
                   2.184755575062588397026324600\\
                  83&  3&         4&   1& {\rm m}034(-3,2) &
                   2.207666238726932912474919817\\
              215811&  6&         6&   1& {\rm m}015(-3,2) &
                   2.226717903919389683617840551\\
             7729991&  7&        32&   1& {\rm m}038(+1,2) &
                   2.259767132595975572056728360\\
                5783&  5&         1&   1& {\rm m}016(+2,3) &
                   2.272631863586558174554150421\\
                1588&  4&         8&   1& {\rm m}038(+4,1) &
                   2.277959444936926552279329057\\
                 275&  4&         2&   5& {\rm m}016(-4,3) &
                   2.343017136901306265356245324\\
                  23&  3&         2&   5& {\rm m}019(+1,4) &
                   2.356768406942319302303249007\\
                2151&  4&        12&   1& {\rm m}015(+8,1) &
                   2.362700792554500476496595823\\
                  31&  3&         2&   3& {\rm m}149(+1,2) &
                   2.374749990937219254249043277\\
                 688&  4&         2&   1& {\rm m}019(+3,4) &
                   2.425585378666368783529163526\\
                 491&  4&         1&   1& {\rm m}029(-3,2) &
                   2.468232196680908678928523005\\
           202734487&  8&       100&   1& {\rm m}038(+3,2) &
                   2.502659305372821115596395708\\
\hline\end{array}$$

\newpage

\begin{center}
{\bf Table~13b:}\quad Rational relations of Dedekind zeta values to volumes
\end{center}
$$\begin{array}{|rr|rr|r|r|}\hline-D&n&a&b&
{\cal M}\qquad{}&{\rm vol}({\cal M})\qquad\qquad\qquad{}\\\hline
                 331&  4&         1&   2& {\rm m}070(-3,1) &
                   2.527418477316087311769432693\\
                  59&  3&         2&   1& {\rm m}039(+6,1) &
                   2.568970600936708884920674169\\
                 507&  4&         1&   1& {\rm m}160(-3,2) &
                   2.595387593686742138301993834\\
            32775179&  7&       316&   1& {\rm m}026(-5,2) &
                   2.609181239513033362390193601\\
             4795631&  7&        13&   1& {\rm m}032(+5,2) &
                   2.629405395288398722278830854\\
                4903&  5&         2&   3& {\rm m}034(-5,2) &
                   2.635689043782677044312119897\\
                   7&  2&         4&   1& {\rm m}036(-4,3) &
                   2.666744783449059790796712462\\
            58360112&  7&       788&   1& {\rm m}034(+5,2) &
                   2.679475805755312597797042568\\
                 751&  4&         2&   1& {\rm m}081(-4,1) &
                   2.781833912396079791875337802\\
                1156&  4&         4&   1& {\rm m}082(+1,3) &
                   2.786804556415568521855509429\\
            12558899&  7&        68&   1& {\rm m}070(-3,2) &
                   2.812516496543210373175617462\\
                  23&  3&         1&   3& {\rm m}221(+3,1) &
                   2.828122088330783162763898809\\
             4297259&  7&        10&   1& {\rm m}221(-1,2) &
                   2.913332114306066935687193484\\
                1192&  4&         4&   1& {\rm m}148(-3,2) &
                   2.921511428929383082004060113\\
                 283&  4&         1&   3& {\rm m}130(-2,3) &
                   2.944106486676696264274356569\\
                 848&  4&         2&   1& {\rm m}207(-1,3) &
                   2.958372867591394347636130075\\
               60803&  5&        28&   1& {\rm m}148(+5,1) &
                   2.970321110188936428725353762\\
                   3&  2&         2&   3& {\rm m}149(-4,1) &
                   3.044824819228960875063607662\\
             3998639&  7&        17&   2& {\rm m}286(-4,1) &
                   3.058954658860052525649857259\\
                 563&  4&         1&   1& {\rm m}130(-4,1) &
                   3.059338057778955673338809625\\
                4511&  5&         1&   2& {\rm m}119(+3,2) &
                   3.087137822943710148656945886\\
                7031&  5&         1&   1& {\rm m}160(-4,1) &
                   3.104808522680010091051472945\\
          8843652791&  9&       571&   1& {\rm m}115(-5,2) &
                   3.123273828139318565185117904\\
               58064&  5&        40&   1& {\rm m}141(+2,3) &
                   3.133349648660896351371796912\\
                 331&  4&         2&   5& {\rm m}146(+5,1) &
                   3.159273096645109139711790867\\
                  31&  3&         1&   2& {\rm s}119(+4,1) &
                   3.166333321249625672332057703\\
                  44&  3&         1&   1& {\rm m}141(+4,1) &
                   3.177293278600325976353827624\\
               11243&  5&         2&   1& {\rm s}090(+5,1) &
                   3.252908048471645923807355063\\
                 107&  3&         4&   1& {\rm m}168(-3,2) &
                   3.275871643943933942369560370\\
             3685907&  7&         7&   1& {\rm m}222(+3,2) &
                   3.299219431617328241596791763\\
                  23&  3&         2&   7& {\rm m}178(-2,3) &
                   3.299475769719247023224548610\\
             1290496&  6&        80&   1& {\rm m}285(-4,1) &
                   3.341002200879537591226767603\\
              463471&  6&        16&   1& {\rm m}249(+4,1) &
                   3.362093204427048043707589278\\
              899447&  6&        44&   1& {\rm m}189(+3,2) &
                   3.383197893650556120356335305\\
                7463&  5&         1&   1& {\rm m}178(+4,3) &
                   3.402991251166455752574894719\\
                5783&  5&         2&   3& {\rm m}175(-1,3) &
                   3.408947795379837261831225632\\
                2068&  4&         8&   1& {\rm m}389(+3,1) &
                   3.410187936572092377001210125\\
                 283&  4&         2&   7& {\rm m}189(-5,2) &
                   3.434790901122812308320082664\\
                 976&  4&         2&   1& {\rm m}286(-5,1) &
                   3.454313917492031906374420374\\
              753079&  6&        26&   1& {\rm m}337(-3,1) &
                   3.474247761312742296029008553\\
\hline\end{array}$$

\newpage

\begin{center}
{\bf Table~13c:}\quad Rational relations of Dedekind zeta values to volumes
\end{center}
$$\begin{array}{|rr|rr|r|r|}\hline-D&n&a&b&
{\cal M}\qquad{}&{\rm vol}({\cal M})\qquad\qquad\qquad{}\\\hline
               23103&  5&         6&   1& {\rm m}223(+5,1) &
                   3.476375673391812878562982302\\
                4903&  5&         1&   2& {\rm m}220(+5,2) &
                   3.514252058376902725749493187\\
               11551&  5&         2&   1& {\rm m}223(-1,3) &
                   3.544081734644579868884505442\\
             2803244&  6&       236&   1& {\rm m}192(-5,2) &
                   3.550430141181664375652339763\\
            81589747&  7&       828&   1& {\rm m}213(-5,2) &
                   3.573600148051412384835850940\\
                 643&  4&         1&   1& {\rm m}247(-1,4) &
                   3.581707325568365305142272152\\
               92779&  6&         1&   1& {\rm m}222(-6,1) &
                   3.608689061770784943522497291\\
                   4&  2&         1&   1& {\rm s}942(-2,1) &
                   3.663862376708876060218414059\\
                1399&  4&         4&   1& {\rm m}293(+2,3) &
                   3.675645605949870731162818740\\
               29963&  5&         8&   1& {\rm m}310(+1,2) &
                   3.687171944653355877440909509\\
               94363&  6&         1&   1& {\rm m}345(+1,2) &
                   3.702897321856940612616393443\\
             1014119&  6&        38&   1& {\rm m}286(-6,1) &
                   3.719977654342577711072823353\\
                 104&  3&         4&   1& {\rm s}297(+1,3) &
                   3.758844948237284271433258100\\
                  23&  3&         1&   4& {\rm s}645(-2,1) &
                   3.770829451107710883685198412\\
                 331&  4&         1&   3& {\rm s}254(+5,1) &
                   3.791127715974130967654149040\\
                  59&  3&         4&   3& {\rm s}296(-5,1) &
                   3.853455901405063327381011254\\
               17348&  5&         4&   1& {\rm s}403(-1,2) &
                   3.883006168054935587460640255\\
                 283&  4&         1&   4& {\rm m}339(-2,3) &
                   3.925475315568928352365808759\\
              112919&  5&        88&   1& {\rm m}304(+1,3) &
                   3.933950637784033249426550295\\
                  76&  3&         2&   1& {\rm s}784(+1,2) &
                   3.970289623890655394010469558\\
               12447&  5&         2&   1& {\rm s}657(-1,2) &
                   3.978127852359131526367873361\\
               38083&  5&        14&   1& {\rm s}437(+1,3) &
                   4.003979154528882088183782858\\
                1099&  4&         2&   1& {\rm s}663(+1,2) &
                   4.018817238361670351502527938\\
               10407&  5&         3&   2& {\rm m}322(+3,2) &
                   4.028673167553685008557652954\\
                6724&  4&        64&   1& {\rm s}479(-5,1) &
                   4.043986894313186522280561166\\
        271488204251& 10&      3669&   2& {\rm s}394(+5,2) &
                   4.057706182949845691594135513\\
                   3&  2&         1&   2& {\rm s}912(+0,1) &
                   4.059766425638614500084810217\\
              629952&  6&        18&   1& {\rm s}386(+5,2) &
                   4.061101845242785944343938727\\
                 448&  4&         1&   2& {\rm s}566(+2,3) &
                   4.116968736386066724912101479\\
             1107052&  6&        50&   1&{\rm v}1315(-4,1) &
                   4.156675426817942891730738689\\
               13219&  5&         2&   1& {\rm s}645(+1,3) &
                   4.171320401322500074447402835\\
                 751&  4&         4&   3& {\rm s}850(-3,1) &
                   4.172750868594119687813006703\\
            66467451&  7&       480&   1& {\rm s}523(-6,1) &
                   4.201030415031789207347923442\\
                5519&  5&         1&   2& {\rm s}648(+5,1) &
                   4.229135386220444476180426062\\
              161939&  6&         2&   1& {\rm s}682(+3,1) &
                   4.230216834783897300285746784\\
                  23&  3&         2&   9& {\rm s}702(-3,1) &
                   4.242183132496174744145848213\\
                  87&  3&         2&   1& {\rm s}784(-1,2) &
                   4.252582946954347612792724802\\
             1825672&  6&       100&   1& {\rm m}358(-5,3) &
                   4.259629131248291580063933525\\
                8647&  5&         1&   1&{\rm v}0940(-5,2) &
                   4.268032673602804301572090960\\
                 400&  4&         2&   5& {\rm m}400(+4,1) &
                   4.306207600730808652919837159\\
\hline\end{array}$$

\newpage

\begin{center}
{\bf Table~13d:}\quad Rational relations of Dedekind zeta values to volumes
\end{center}
$$\begin{array}{|rr|rr|r|r|}\hline-D&n&a&b&
{\cal M}\qquad{}&{\rm vol}({\cal M})\qquad\qquad\qquad{}\\\hline
              365263&  6&        26&   3& {\rm s}961(+1,2) &
                   4.322097020182094625847110669\\
              104483&  6&         1&   1& {\rm s}648(+1,2) &
                   4.330099508377546093234851467\\
              238507&  6&         7&   2& {\rm s}649(-4,1) &
                   4.369511150125176794052649201\\
             4297259&  7&        20&   3& {\rm s}730(-1,2) &
                   4.369998171459100403530790226\\
               29444&  5&         8&   1& {\rm s}478(-1,3) &
                   4.374966511983605191446145441\\
                1968&  4&         6&   1& {\rm s}594(+1,3) &
                   4.403155016694858100998436421\\
                  83&  3&         2&   1& {\rm s}869(-1,2) &
                   4.415332477453865824949839635\\
                 283&  4&         2&   9& {\rm s}650(-1,3) &
                   4.416159730015044396411534854\\
                 331&  4&         2&   7& {\rm m}400(+4,3) &
                   4.422982335303152795596507213\\
            50052727&  7&       296&   1&{\rm v}2200(-3,2) &
                   4.454453084485924995476446657\\
                 116&  3&         4&   1& {\rm s}881(-1,3) &
                   4.464658911548680772053932692\\
             7729991&  7&        16&   1& {\rm s}645(+4,3) &
                   4.519534265191951144113456721\\
                5783&  5&         1&   2&{\rm v}2203(+3,1) &
                   4.545263727173116349108300843\\
                1588&  4&         4&   1&{\rm v}2641(-4,1) &
                   4.555918889873853104558658114\\
               79952&  5&        36&   1& {\rm s}594(+3,2) &
                   4.606469152444949889745752026\\
               22331&  5&         4&   1& {\rm s}884(+2,3) &
                   4.611024261895638050709808503\\
                 731&  4&         1&   1& {\rm s}649(-3,4) &
                   4.626565091277539615466468128\\
                4511&  5&         1&   3& {\rm s}646(+5,2) &
                   4.630706734415565222985418829\\
            75117248&  7&       616&   1&{\rm v}1788(+3,2) &
                   4.646329951144689812397743785\\
                1107&  4&         2&   1& {\rm s}928(+4,1) &
                   4.662289290947371076825746909\\
         99961920379&  9&     15436&   1& {\rm s}649(-5,3) &
                   4.678743072215143322016171423\\
              141791&  5&       104&   1& {\rm s}707(+5,1) &
                   4.682218629386119691937245602\\
                 275&  4&         1&   5&{\rm v}1251(+4,3) &
                   4.686034273802612530712490649\\
                  23&  3&         1&   5& {\rm s}944(-1,2) &
                   4.713536813884638604606498015\\
                2151&  4&         6&   1& {\rm s}882(+4,1) &
                   4.725401585109000952993191647\\
                  31&  3&         1&   3& {\rm s}874(+4,1) &
                   4.749499981874438508498086555\\
                  44&  3&         2&   3&{\rm v}1368(+2,3) &
                   4.765939917900488964530741437\\
          2095218667&  8&      2240&   1& {\rm s}901(-3,2) &
                   4.809367033602469652109958201\\
               14103&  5&         2&   1& {\rm s}784(+5,2) &
                   4.814768023880028928726297311\\
                 688&  4&         1&   1& {\rm s}944(-3,2) &
                   4.851170757332737567058327052\\
               31684&  5&         8&   1&{\rm v}2381(+3,1) &
                   4.868856851098063200826956663\\
                1879&  4&         4&   1&{\rm v}2914(+2,3) &
                   4.875758159106157239356295909\\
               11243&  5&         4&   3&{\rm v}2447(+1,3) &
                   4.879362072707468885711032595\\
             9429911&  7&        20&   1& {\rm s}838(-2,3) &
                   4.883386971539476550563987375\\
                 283&  4&         1&   5&{\rm v}3215(+3,1) &
                   4.906844144461160440457260948\\
              239639&  5&       184&   1& {\rm s}918(+3,2) &
                   4.924074751099801277306001435\\
                 491&  4&         1&   2& {\rm s}900(-1,3) &
                   4.936464393361817357857046010\\
             3685907&  7&        14&   3&{\rm v}2335(+3,2) &
                   4.948829147425992362395187645\\
             6515927&  7&        11&   1& {\rm s}900(+3,2) &
                   4.953010368136679742472918540\\
               23339&  5&         4&   1&{\rm v}3199(+3,1) &
                   4.967241778215442305200322817\\
\hline\end{array}$$

\newpage

\begin{center}
{\bf Table~13e:}\quad Rational relations of Dedekind zeta values to volumes
\end{center}
$$\begin{array}{|rr|rr|r|r|}\hline-D&n&a&b&
{\cal M}\qquad{}&{\rm vol}({\cal M})\qquad\qquad\qquad{}\\\hline
             5873596&  6&       688&   1&{\rm v}1858(+6,1) &
                   4.988809551743365540758322561\\
           948381887&  8&       496&   1&{\rm v}2486(-3,2) &
                   4.989804907785885826671653092\\
             7557047&  7&        14&   1&{\rm v}2221(-1,3) &
                   5.002053292789971186622200159\\
           202734487&  8&        50&   1& {\rm s}900(+1,3) &
                   5.005318610745642231192791416\\
                 331&  4&         1&   4&{\rm v}2402(+3,2) &
                   5.054836954632174623538865387\\
          2401259831&  8&      1978&   1&{\rm v}3184(+4,1) &
                   5.056718039142583479359939301\\
                   3&  2&         2&   5&{\rm v}2422(+1,3) &
                   5.074708032048268125106012771\\
                7463&  5&         2&   3&{\rm v}2344(+5,2) &
                   5.104486876749683628862342079\\
       7748687650003& 10&    232080&   1& {\rm s}900(+2,3) &
                   5.109227541548022851884464403\\
               14631&  5&         2&   1& {\rm s}958(+3,2) &
                   5.112724558199808581711904721\\
                  59&  3&         1&   1&{\rm v}3066(+1,2) &
                   5.137941201873417769841348339\\
            43210364&  7&       242&   1&{\rm v}3305(-1,2) &
                   5.172428768697906186835081520\\
              709783&  6&        20&   1& {\rm s}952(-4,1) &
                   5.175230582596977610906843350\\
                 507&  4&         1&   2&{\rm v}3347(+3,1) &
                   5.190775187373484276603987668\\
   12476239474594496& 12&   9408656&   1&{\rm v}2824(+4,1) &
                   5.194214571520112044514895549\\
                 139&  3&         4&   1&{\rm v}3106(+3,1) &
                   5.198433660442561125403410328\\
                9439&  5&         1&   1&{\rm v}2759(-3,1) &
                   5.200723713644593903398311904\\
                1732&  4&         4&   1&{\rm v}3187(-4,1) &
                   5.202496842480823311129029464\\
            32775179&  7&       158&   1&{\rm v}2725(-4,1) &
                   5.218362479026066724780387202\\
            46692071&  7&       250&   1&{\rm v}2825(-4,1) &
                   5.240170302454763905979008831\\
               14911&  5&         2&   1&{\rm v}2704(-5,1) &
                   5.262436101311908292112209718\\
                4903&  5&         1&   3& {\rm s}944(-5,2) &
                   5.271378087565354088624239780\\
               13523&  5&         2&   1&{\rm v}2530(+1,3) &
                   5.287936270526127135612285367\\
               70736&  5&        32&   1&{\rm v}2787(-1,3) &
                   5.288937507218637514049542257\\
                   7&  2&         2&   1&{\rm v}3390(+3,1) &
                   5.333489566898119581593424925\\
            58360112&  7&       394&   1&{\rm v}3462(-1,2) &
                   5.358951611510625195594085136\\
                1791&  4&         4&   1& {\rm s}961(+2,3) &
                   5.363693221795981744185777767\\
             7792864&  6&       976&   1&{\rm v}2789(-2,3) &
                   5.387253764656890006255173367\\
               34779&  5&         9&   1&{\rm v}3214(+3,1) &
                   5.426764227123098991537817177\\
                9759&  5&         1&   1&{\rm v}3031(+3,1) &
                   5.473032966608735883824871653\\
                   4&  2&         2&   3&{\rm v}3412(+5,1) &
                   5.495793565063314090327621089\\
               13883&  5&         2&   1&{\rm v}3310(+5,1) &
                   5.504748837818013496725737495\\
               34436&  5&         8&   1&{\rm v}2794(-2,3) &
                   5.509676989525577537121576098\\
             4241707&  6&       296&   1&{\rm v}3428(-4,1) &
                   5.517978522012509993351570954\\
                  31&  3&         2&   7&{\rm v}3091(-2,3) &
                   5.541083312186844926581100981\\
                 751&  4&         1&   1&{\rm v}3277(-2,3) &
                   5.563667824792159583750675604\\
                1156&  4&         2&   1&{\rm v}3183(-3,2) &
                   5.573609112831137043711018858\\
              688927&  6&        14&   1&{\rm v}3243(-3,1) &
                   5.576259626360039431093360827\\
            12558899&  7&        34&   1&{\rm v}3520(+4,1) &
                   5.625032993086420746351234925\\
         48502810352&  9&      5230&   1&{\rm v}3157(+5,1) &
                   5.646678958454479680102882510\\
\hline\end{array}$$

\newpage

\begin{center}
{\bf Table~13f:}\quad Rational relations of Dedekind zeta values to volumes
\end{center}
$$\begin{array}{|rr|rr|r|r|}\hline-D&n&a&b&
{\cal M}\qquad{}&{\rm vol}({\cal M})\qquad\qquad\qquad{}\\\hline
             7729991&  7&        64&   5&{\rm v}3264(+4,1) &
                   5.649417831489938930141820901\\
                  23&  3&         1&   6&{\rm v}3375(-3,2) &
                   5.656244176661566325527797618\\
                2319&  4&         6&   1&{\rm v}2944(-5,2) &
                   5.671883543566564169608496809\\
         81051965432&  8&    636704&   1&{\rm v}3107(+2,3) &
                   5.683882556138987145289674657\\
                 331&  4&         2&   9&{\rm v}3438(-3,1) &
                   5.686691573961196451481223560\\
             1494223&  6&        56&   1&{\rm v}3036(+3,2) &
                   5.687282054652596214890785149\\
                3312&  4&        12&   1&{\rm v}3209(+2,3) &
                   5.724388054820026594606305363\\
            97569124&  7&       736&   1&{\rm v}3214(+2,3) &
                   5.736795047142670277771627006\\
               22424&  5&         4&   1&{\rm v}3246(-2,3) &
                   5.745000104449036634165968002\\
              365263&  6&        13&   2&{\rm v}3184(-3,2) &
                   5.762796026909459501129480892\\
              561863&  6&        10&   1&{\rm v}3239(+3,2) &
                   5.804174400001677001746529114\\
                1192&  4&         2&   1&{\rm v}3214(-4,3) &
                   5.843022857858766164008120227\\
                 283&  4&         1&   6&{\rm v}3431(-2,3) &
                   5.888212973353392528548713138\\
                 848&  4&         1&   1&{\rm v}3477(+4,1) &
                   5.916745735182788695272260151\\
              661831&  6&        16&   1&{\rm v}3361(+1,3) &
                   5.920105898675782923200528147\\
                1927&  4&         4&   1&{\rm v}3452(-5,1) &
                   5.934463883899472497426225266\\
              417467&  6&        13&   2&{\rm v}3375(-5,2) &
                   5.999880841314256220228307651\\
               10407&  5&         1&   1&{\rm v}3418(+6,1) &
                   6.043009751330527512836479431\\
                1371&  4&         2&   1&{\rm v}3489(+2,3) &
                   6.087435457969104359353387839\\
                   3&  2&         1&   3&{\rm v}3492(-4,1) &
                   6.089649638457921750127215325\\
              107264&  5&        52&   1&{\rm v}3454(-5,1) &
                   6.120178528987339574492892096\\
                1255&  4&         2&   1&{\rm v}3492(+4,3) &
                   6.223431719518907660423581730\\
               13219&  5&         4&   3&{\rm v}3543(+1,3) &
                   6.256980601983750111671104253\\
               58064&  5&        20&   1&{\rm v}3526(+2,3) &
                   6.266699297321792702743593825\\
\hline\end{array}$$

\bet{14}{Further single-complex-place fields, from cusped manifolds}
\tabw{x^2+2}{8}{2}{1}{{\rm v}2787}\hline
\tabw{x^3-x^2-2x-2}{152}{4}{1}{{\rm v}3526}\hline
\tabw{x^4-3x^2-2x+1}{1328}{2}{1}{{\rm v}2631}
\tabw{x^4-x^3+x^2-6x-4}{2375}{4}{1}{{\rm v}3545}
\tabw{x^4+5x^2-3}{4107}{15}{1}{{\rm v}1143}\hline
\tabw{x^5-2x^4+3x^2-2x-1}{7367}{1}{1}{{\rm m}052}\hline
\tabw{x^6-x^5-2x^4-x^3+3x^2+2x-1}{303619}{5}{1}{{\rm s}281}
\tabw{x^6-x^5-2x^4+4x^3-10x^2+6x+3}
{13266363}{1362}{1}{{\rm v}1461}\hline
\tabw{x^7-2x^6-6x^5+9x^4+12x^3-9x^2-11x+2}
{161329612}{1832}{1}{{\rm v}3418}\hline
\tabw{x^8-4x^7+5x^6-x^5-6x^4+9x^3-4x+1}
{74671875}{12}{1}{{\rm m}283}
\tabw{x^8-3x^7-x^6+4x^5+8x^4-4x^3-8x^2+1}
{397538359}{142}{1}{{\rm v}2824}
\tabw{x^8-x^7-6x^6+8x^5-12x^3+23x^2-9x-3}
{2597840403}{2853}{1}{{\rm s}311}
\ent

\newpage

\begin{center}{\bf Table~15:}\quad Joins of quadratics\end{center}
$$\begin{array}{|l|rr|rr|rrr|c|}\hline\mbox{quartic join}
&n_1&-D_1&n_2&-D_2&a&b_1&b_2&\mbox{manifolds}\\\hline
x^4-x^2+1&2&3&2&4&2&2&1&{\rm m}350(-1,3)\\&&&&&
&&&{\rm m}360(-2,3)\\&&&&&
1&1&1&{\rm s}913,{\rm v}2274\\\hline
x^4-x^3-x^2-2x+4&2&3&2&7&4&4&1&{\rm v}2274(\pm4,1)\\\hline
x^4-3x^2+4&2&4&2&7&4&2&1&{\rm v}1859(\pm3,1)\\&&&&&
&&&{\rm v}1859(\pm1,3)\\&&&&&
&&&{\rm m}314\mbox{--}5\\\hline
\end{array}$$

\begin{center}{\bf Table~16:}\quad Joins of a quadratic and cubic\end{center}
$$\begin{array}{|l|rr|rr|rrr|c|}\hline\mbox{sextic join}
&n_1&-D_1&n_2&-D_2&a&b_1&b_2&\mbox{manifolds}\\\hline
x^6-x^5+x^4-2x^3+x^2+1&2&3&3&23&1&1&3&{\rm v}2274(\pm3,2)\\\hline
x^6-x^5-3x^3+2x^2+x+1&2&3&3&44&1&1&1&{\rm v}2274(\pm6,1)\\&&&&&
&&&{\rm v}2274(\pm2,3)\\\hline
x^6-2x^4-2x^3+4x^2+2x+1&2&3&3&59&2&2&1&{\rm v}2274(\pm1,2)\\&&&&&
&&&{\rm s}636(-1,4)\\&&&&&
&&&{\rm s}618(+1,4)\\\hline
x^6-4x^4+4x^2+1&2&4&3&59&2&1&1&{\rm s}518(-1,4)\\&&&&&
&&&{\rm s}530(-1,4)\\&&&&&
2&2&1&{\rm v}3066\\\hline
x^6-6x^4-5x^3+16x^2+8x+8&2&7&3&59&4&1&2&{\rm v}3066(\pm4,1)\\\hline
\end{array}$$

\begin{center}{\bf Table~17:}\quad Joins of cubics\end{center}
$$\begin{array}{|l|rr|rr|rrr|c|}\hline\mbox{nonadic join}
&n_1&-D_1&n_2&-D_2&a&b_1&b_2&\mbox{manifolds}\\\hline
x^9+2x^7-2x^6+8x^5+4x^4&3&23&3&59&2&6&1&
{\rm v}3066(\pm3,2)\\{}\quad
+11x^3+4x^2-1&&&&&&&&\\\hline
x^9-2x^7-5x^6+12x^5+8x^4&3&44&3&59&2&2&1&
{\rm v}3066(\pm2,3)\\{}\quad
+15x^3+4x^2+2x-1&&&&&&&&{\rm v}3066(\pm6,1)\\\hline
\end{array}$$

\newpage

\begin{center}{\bf Table~18:}\quad Joins of a quadratic
and quartic\end{center}
$$\begin{array}{|l|rr|rr|rrr|c|}\hline\mbox{quartic }\subset
\mbox{ octadic join}
&n_1&-D_1&n_2&-D_2&a&b_1&b_2&\mbox{manifolds}\\\hline
x^4-2x^3+2&2&4&4&400&4&1&5&{\rm m}135(\pm1,4)\\
\subset x^8-2x^7+2x^6+2x^5-2x^4&&&&&
4&3&5&{\rm v}1859(\pm4,1)\\
\qquad{}+2x^3+2x^2-2x+1&&&&&
&&&{\rm v}1859(\pm1,4)\\&&&&&
2&1&5&{\rm v}2942\mbox{--}7\\\hline
x^4-x^3+x+1&2&7&4&448&8&1&8&{\rm m}235(-4,1)\\
\subset x^8+4x^6+x^4-6x^2+4&&&&&
&&&{\rm m}234(-1,3)\\&&&&&
&&&{\rm m}305(-4,1)\\&&&&&
4&1&4&{\rm s}719(+7,1)\\&&&&&
&&&{\rm v}1373(-2,3)\\&&&&&
4&1&8&{\rm v}3505\mbox{--}7\\\hline
x^4-x^3-x^2+x+1&2&3&4&507&2&1&1&{\rm m}023(-5,1)\\
\subset x^8-x^7+2x^6+3x^5-x^4&&&&&
&&&{\rm m}022(+2,3)\\
\qquad{}+3x^3+2x^2-x+1&&&&&
&&&{\rm m}038(-5,1)\\&&&&&
1&1&1&{\rm s}645(-2,3)\\&&&&&
&&&{\rm s}646(-2,3)\\&&&&&
&&&{\rm s}648(-7,1)\\&&&&&
&&&{\rm s}649(+2,3)\\&&&&&
&&&{\rm v}1809(-4,1)\\&&&&&
&&&{\rm m}345\\&&&&&
1&2&1&{\rm v}3461\mbox{--}2\\\hline
x^4+x^2-2x+1&2&4&4&1156&4&4&1&{\rm v}3318\mbox{--}9\\
\subset x^8+5x^6+4x^4+5x^2+1&&&&&&&&\\\hline
x^4-x^3-2x^2+3&2&3&4&4107&30&45&1&{\rm s}869\\
\subset x^8-5x^6+28x^4+15x^2+9&&&&&&&&\\\hline
x^4-2x^3-x^2+2x+2&2&4&4&6724&128&32&1&{\rm m}135(\pm2,3)\\
\subset x^8+13x^6+40x^4+52x^2+16&&&&&
128&96&1&{\rm v}1859(\pm2,3)\\&&&&&
&&&{\rm v}1859(\pm3,2)\\&&&&&
64&32&1&{\rm v}3431(+3,2)\\&&&&&
&&&{\rm v}3217(-5,1)\\&&&&&
&&&{\rm v}3213(-5,1)\\&&&&&
&&&{\rm v}3212(+4,3)\\&&&&&
&&&{\rm v}3209(+4,3)\\&&&&&
&&&{\rm v}3210(+5,1)\\&&&&&
&&&{\rm v}3387(-2,3)\\&&&&&
&&&{\rm v}3207(+5,1)\\&&&&&
&&&{\rm v}3208(+4,3)\\&&&&&
&&&{\rm s}937\mbox{--}41\\&&&&&
&&&{\rm v}2573\mbox{--}6\\\hline
\end{array}$$

\newpage

\begin{center}
{\bf Table~19:}\quad Numerical values of $Z_{|D|}:=Z_K$ for
imaginary-quadratic fields
\end{center}
$$\begin{array}{|r|r|}\hline
-D&Z_{|D|}\qquad\qquad\qquad\qquad\qquad\qquad{}\\\hline
 3 &   2.02988321281930725004240510854904057188337861506059\\
 4 &   3.66386237670887606021841405972953644309659749712668\\
 7 &  10.66697913379623916318684985044260017639353555421055\\
 8 &  12.04609204009437764726837862923359423099605804944499\\
11 &  16.59129969483175048405984013396780188163367504042159\\
15 &  37.66336673357521501108052592233790231511162680252581\\
20 &  50.44763111371256002113427103608556540514680566830894\\
24 &  62.18607477383502595106662726058243112965063718233095\\
39 & 165.57570369926833581678917631121833077904271386139253\\
84 & 404.73628202464445448608555478460494570828840108112927\\
\hline\end{array}$$

\newpage

\begin{center}
{\bf Table~20:}\quad
Volume of the cubical and octahedral links
\end{center}
\begin{verbatim}
            114.5376114512362943224424724789936249435748808242
            20429231250851942121127501027441161281234687892520
            99421046573180631580080289831770300702374092992825
            00845864602839836513300827269407599687597029206605
            31531580759324150754143878837313712948751658431691
            71743093772784113318628123855416344693593219229652
            66073261071127716611194216866092756637015575724347
            30495583461169092818889421389435922005281897761327
            87481780722702810697927326811113193211506865618819
            83951500523880503765622058805105992474582567995845
            24852839534204871794340415879015640464651359451611
            96357246969286008273298748180867981112146197992045
            62818897243423342214341286694709315899967336066828
            40288093494107303594494398978521250331659884338246
            44517641804581874967125904277881596619253538707806
            55344565593252496994635391954711081416596646456178
            99682627682970247100159721685353632474114663301702
            67227010018769548802600661656974059930577678444203
            19782424535454491101271247208720911167848941140775
            80409525758170311502822254452974913073005473499436
            49644923893596219534503892584099468539202882445560
            67043214403394086453561263528712168065777974863536
            20425213800992134976726381671853648499562580833637
            68477163430037441223365797732060301796216060089171
            27284752615537852510660945583180384146439018576726
            91389752575734455792451007575910156390208650947188
            83396726765778146515970887067941025010073059637530
            81422959270328728829842816188973146607449652158522
            78007793547439090335284739905795279447511546662295
            82830239755327066934515195332331267833862837558054
            31913571195670329744639917173868105940089705910151
            24896982801226879473447589093589206619826480243308
            26876906144012908721342540125943215276251702819029
            18946884384227781081283744525707741068411779905750
            35583054992839794034527044721173256021542771843790
            61496448437835045757339558623479750202375305175456299
\end{verbatim}

\newpage

\begin{center}
{\bf Table~21:}\quad
Volume of the dodecahedral and icosahedral links
\end{center}
\begin{verbatim}
            310.9131454258982418156066571340803789107179627546
            95386653927493346989744424412131109973939975971507
            98643269709265823832914978469249739351095948216115
            51135502802407570241514603614987653238948839299669
            45789956365988157798923976701962012526567526838465
            48548800271771303442838923804565772591836525869728
            54084324901367744756506352594813054449587164285970
            71563611411064456886341052373960417941929693593205
            04772124188342369610921968892286063808069013825878
            15230467416868749934315874565595988479080697713326
            50858675650630654788112165890478712971506603197258
            23113347663540660063185505122390690532209442720746
            84653584725667817894996050270455301963078167128747
            88497693605671527599227302568072978145795362466734
            51204820032785401797775839422955000755158363428262
            34566607211990453989775867832671557955314292042596
            40210765494283340093378373891402748722649628433618
            99776582050245619951448951311088720526789005381822
            56286704026182535081549522339738817785994375329737
            22530736120788811969510398747558459417006190535250
            16187420583314157401670875056552682715430880984829
            28187061043663478707803859734383346993815502724978
            25927982584761501331976838167147091050425539561024
            58379417142495796685978885556943637012844570671457
            05212348658659424666682968591727400295475117140581
            88231035672842687990189221264859638768219728313394
            01215816266262935688361101821397595969743744268728
            61897426110723765254305981655479282036799500110828
            57792554522835448609248318052260233556335883617840
            74552968702184547644390948427990004868424690548734
            32941297657691799943811633733468108748967009155145
            80688308119967508389482981267982511015711538776266
            89088501076905467164741221851557035108445066926964
            88328174086883755702372138688832584538642639144251
            47198193809500775033230947560721000905153519571355423
\end{verbatim}

\newpage

\begin{center}
{\bf Table~22:}\quad
Empirical reductions of Feynman orthoschemes to the values of Table~19
\end{center}
$$\begin{array}{|ccc|c||ccc|c||ccc|c|}\hline
\psi_1&\psi_2&\psi_3&S(\psi_1,\psi_2,\psi_3)&
\psi_1&\psi_2&\psi_3&S(\psi_1,\psi_2,\psi_3)&
\psi_1&\psi_2&\psi_3&S(\psi_1,\psi_2,\psi_3)\\[3pt]\hline
0&0&0&                                       Z_{4}&
0&0&{\pi\over6}&                        \df53Z_{3}&
0&0&{\pi\over4}&                        \df14Z_{8}\\[3pt]
0&0&{\pi\over3}&                        \df23Z_{4}&
0&{\pi\over6}&0&                        \df53Z_{3}&
0&{\pi\over6}&{\pi\over6}&              \df56Z_{4}\\[3pt]
0&{\pi\over6}&{\pi\over4}&           \df1{24}Z_{24}&
0&{\pi\over6}&{\pi\over3}&              \df56Z_{3}&
0&{\pi\over4}&0&                        \df14Z_{8}\\[3pt]
0&{\pi\over4}&{\pi\over6}&           \df1{24}Z_{24}&
0&{\pi\over4}&{\pi\over4}&              \df12Z_{4}&
0&{\pi\over4}&{\pi\over3}&         \df{1}{12}Z_{8}\\[3pt]
0&{\pi\over3}&0&                        \df23Z_{4}&
0&{\pi\over3}&{\pi\over6}&              \df56Z_{3}&
0&{\pi\over3}&{\pi\over4}&           \df1{12}Z_{8}\\[3pt]
0&{\pi\over3}&{\pi\over3}&              \df16Z_{4}&
{\pi\over6}&0&{\pi\over6}&           \df1{12}Z_{15}&
{\pi\over6}&0&{\pi\over4}&          \df1{144}Z_{84}\\[3pt]
{\pi\over6}&0&{\pi\over3}&           \df1{72}Z_{39}&
{\pi\over6}&{\pi\over6}&{\pi\over6}&    \df16Z_{11}&
{\pi\over6}&{\pi\over6}&{\pi\over4}& \df1{16}Z_{15}\\[3pt]
{\pi\over6}&{\pi\over6}&{\pi\over3}&    \df34Z_{3}&
{\pi\over6}&{\pi\over4}&{\pi\over6}& \df5{24}Z_{7}&
{\pi\over6}&{\pi\over4}&{\pi\over4}& \df5{12}Z_{4}\\[3pt]
{\pi\over6}&{\pi\over4}&{\pi\over3}& \df1{48}Z_{15}&
{\pi\over6}&{\pi\over3}&{\pi\over6}&    \df12Z_{3}&
{\pi\over6}&{\pi\over3}&{\pi\over4}& \df5{24}Z_{3}\\[3pt]
{\pi\over6}&{\pi\over3}&{\pi\over3}& \df1{12}Z_{3}&
{\pi\over4}&0&{\pi\over4}&              \df54Z_{3}&
{\pi\over4}&0&{\pi\over3}&           \df1{24}Z_{20}\\[3pt]
{\pi\over4}&{\pi\over6}&{\pi\over4}&    \df16Z_{8}&
{\pi\over4}&{\pi\over6}&{\pi\over3}&    \df58Z_{3}&
{\pi\over4}&{\pi\over4}&{\pi\over4}&    \df14Z_{4}\\[3pt]
{\pi\over4}&{\pi\over4}&{\pi\over3}& \df1{12}Z_{4}&
{\pi\over3}&0&{\pi\over3}&              \df16Z_{7}&
{\pi\over3}&{\pi\over6}&{\pi\over3}&    \df13Z_{3}\\[3pt]
\hline\end{array}$$

\begin{center}{\bf Table~23:}\quad Maximally symmetric knots
to 10 crossings\end{center}
$$\begin{array}{|ll|l|l|r|r|rr|}\hline
\mbox{Rolfsen}&\mbox{HTW}&\mbox{sym}&\mbox{invariant trace field}&
\mbox{sig}&D&a&b\\\hline
4_1&{\rm a}4.1&D_4&x^2-x+1&[0,1]&-3&1&1\\\hline
5_2&{\rm a}5.1&D_2&x^3-x^2+1&[1,1]&-23&1&3\\\hline
6_3&{\rm a}6.1&D_4&x^6-x^5-x^4+2x^3-x+1&[0,3]&-10571&\times&\times\\\hline
7_4&{\rm a}7.6&D_4&x^3+2x-1&[1,1]&-59&1&1\\
7_7&{\rm a}7.1&D_4&x^4+x^2-x+1&[0,2]&257&\times&\times\\\hline
8_{18}&{\rm a}8.12&D_8&x^4-2x^3+x^2-2x+1&[2,1]&-448&1&6\\
8_{21}&{\rm n}8.2&D_2&x^4-x^3+x+1&[0,2]&392&\times&\times\\\hline
9_{35}&{\rm a}9.40&D_6&x^3-2x-2&[1,1]&-76&1&1\\
9_{40}&{\rm a}9.37&D_6&x^4+2x^2-2x+1&[0,2]&592&\times&\times\\
9_{48}&{\rm n}9.6&D_6&x^3-x^2+x+1&[1,1]&-44&1&3\\\hline
10_{123}&{\rm a}10.121&D_{10}&x^4-x^3+x^2-x+1&[0,2]&125&\times&\times\\
10_{157}&{\rm n}10.42&D_4&x^3+x-1&[1,1]&-31&1&8\\\hline
\end{array}$$

\begin{center}{\bf Table~24:}\quad Less symmetric non-alternating
Dedekind-zeta knots\end{center}
$$\begin{array}{|ll|l|l|r|r|rr|}\hline
\mbox{Rolfsen}&\mbox{HTW}&\mbox{sym}&\mbox{invariant trace field}&
\mbox{sig}&D&a&b\\\hline
9_{49}&{\rm n}9.8&D_3&x^3-x^2+1&[1,1]&-23&1&10\\\hline
10_{139}&{\rm n}10.27&D_2&x^4-2x-1&[2,1]&-688&1&1\\
10_{152}&{\rm n}10.36&Z_2&x^5-3x^3-2x^2+2x+1&[3,1]&-8647&1&2\\
10_{153}&{\rm n}10.10&\mbox{triv.}&
x^5-2x^4-2x^3+4x^2-x+1&[3,1]&-29963&4&1\\\hline
\end{array}$$

\newpage

\begin{center}{\bf Table~25:}\quad Maximally symmetric knots
from 11 to 16 crossings\end{center}
$$\begin{array}{|l|l|l|r|}\hline
\mbox{HTW knot}&\mbox{sym}&\mbox{invariant trace field}&
\mbox{sig}\\\hline
{\rm a}11.366&D_6&x^6-x^5+4x^4-3x^3+4x^2-2x-1&[2,2]\\
{\rm n}11.126&D_3&x^5-x^4+3x^2-2x+1&[1,2]\\
{\rm n}11.133&D_3&x^5-x^4+x^2-x+1&[1,2]\\\hline
{\rm a}12.503&D_6&x^7+2x^5+2x^3-x^2-x-1&[1,3]\\
{\rm a}12.561&D_6&x^7-2x^5-x^4+2x^3+x^2+x-1&[1,3]\\
{\rm a}12.1019&D_6&x^{10}-2x^9+2x^8-2x^7-3x^6+7x^5&\\&&
         \qquad{}-3x^4-2x^3+2x^2-2x+1&[2,4]\\
{\rm a}12.1202&D_6&x^8-x^7+x^6+3x^4+x^2-x+1&[0,4]\\
{\rm n}12.555&D_8&x^5+x^3-x^2+2x+1&[1,2]\\
{\rm n}12.642&D_8&x^3-x^2+x+1&[1,1]\\\hline
{\rm a}13.1786&D_8&x^8+2x^6-x^5+3x^4-x^3+x^2-2x+1&[0,4]\\
{\rm a}13.4877&D_8&x^6+4x^4-x^3+4x^2-3x-1&[2,2]\\
{\rm n}13.4051&D_6&x^5-x^4-x^3+x^2+3x+1&[1,2]\\\hline
{\rm a}14.19470&D_{14}&x^6-2x^5+2x^4-3x^3+2x^2-2x+1&[2,2]\\
{\rm n}14.13191&D_5&x^4-x^3-2x+1&[2,1]\\
{\rm n}14.17159&D_5&x^6-2x^5+x^4+x^3+x^2-2x-1&[2,2]\\\hline
{\rm a}15.84903&D_{10}&x^8-2x^7+x^6-2x^5+x^4-4x^3 +6x^2-4x+4&[2,3]\\
{\rm a}15.85262&D_{10}&x^6-3x^4-4x^3+x^2+6x+4&[2,2]\\
{\rm n}15.99226&D_{10}&x^7-x^6+2x^5-x^4+3x^3-3x^2+2x+1&[1,3]\\
{\rm n}15.112310&D_{10}&x^4-2x^3+x^2-2x+1&[2,1]\\\hline
{\rm a}16.379778&D_{16}&x^8-4x^7+6x^6-11x^4+16x^3-10x^2+4x-1&[2,3]\\
{\rm n}16.1007813&D_{9}&x^3+x-1&[1,1]\\\hline
\end{array}$$

\newpage

\setlength{\unitlength}{0.01cm}
\newbox\shell
\newcommand{\dia}[1]{\setbox\shell=\hbox{\begin{picture}(500,800)(-250,-400)#1
\end{picture}}\dimen0=\ht
\shell\multiply\dimen0by7\divide\dimen0by16\raise-\dimen0\box\shell\hfill}

\begin{center}
{\bf Fig.~1:}\quad Alternating platonic link
from light-by-light scattering
\end{center}
\mbox{\hspace{1cm}}\hfill
\dia{
\put(0,-10){\circle{200}}
\put(0,300){\circle{200}}
\put(-280,-240){\circle{200}}
\put( 280,-240){\circle{200}}
\put(-30,270){\line(0,-1){250}}
\put( 30,270){\line(0,-1){250}}
\put(-30,270){\line(1, 0){60}}
\put(-30, 20){\line(1, 0){60}}
\put(-250,-270){\line( 1, 1){240}}
\put(-290,-230){\line( 1,-1){40}}
\put(-290,-230){\line( 1, 1){240}}
\put( -50,  10){\line( 1,-1){40}}
\put( 250,-270){\line(-1, 1){240}}
\put( 290,-230){\line(-1,-1){40}}
\put( 290,-230){\line(-1, 1){240}}
\put(  50,  10){\line(-1,-1){40}}
\put(-310,-230){\line( 1, 2){280}}
\put(-310,-230){\line(-2, 1){60}}
\put(-370,-200){\line( 1, 2){280}}
\put( -90, 360){\line( 2,-1){60}}
\put( 310,-230){\line(-1, 2){280}}
\put( 310,-230){\line( 2, 1){60}}
\put( 370,-200){\line(-1, 2){280}}
\put(  90, 360){\line(-2,-1){60}}
\put(-300,-280){\line( 1, 0){600}}
\put(-300,-340){\line( 1, 0){600}}
\put(-300,-280){\line( 0,-1){60}}
\put( 300,-280){\line( 0,-1){60}}}
\mbox{\hspace{1cm}}

\vfill

\begin{center}
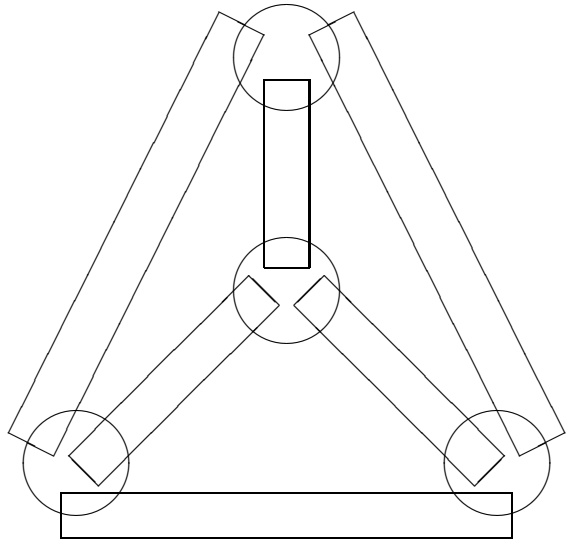
{\bf Fig.~2:}\quad Non-alternating daisy-chain link
\end{center}
\mbox{\hspace{1cm}}\hfill
\dia{
\put(290,78){\circle{200}}
\put(260,150){\circle{200}}
\put(212,212){\circle{200}}
\put(150,260){\circle{200}}
\put(78,290){\circle{200}}
\put(0,300){\circle{200}}
\put(-78,290){\circle{200}}
\put(-150,260){\circle{200}}
\put(-212,212){\circle{200}}
\put(-260,150){\circle{200}}
\put(-290,78){\circle{200}}
\put(-300,0){\circle{200}}
\put(-290,-78){\circle{200}}
\put(-260,-150){\circle{200}}
\put(-212,-212){\circle{200}}
\put(-150,-260){\circle{200}}
\put(-78,-290){\circle{200}}
\put(0,-300){\circle{200}}
\put(78,-290){\circle{200}}
\put(150,-260){\circle{200}}
\put(212,-212){\circle{200}}
\put(260,-150){\circle{200}}
\put(290,-78){\circle{200}}
\put(300,0){\circle{200}}
}
\mbox{\hspace{1cm}}

\newpage

\begin{center}
{\bf Fig.~3:}\quad Alternating figure-8 knot at $D=-3$
\end{center}
\mbox{\hspace{1cm}}\hfill
\dia{
\put(+100,+300){\line(-1,-1){200}}
\put(-100,+300){\line(+1,-1){200}}
\put(-300,+100){\line(+1,-1){300}}
\put(-300,+100){\line(+1,+1){200}}
\put(+300,+100){\line(-1,-1){300}}
\put(+300,+100){\line(-1,+1){200}}
\put(+100,+100){\line(-1,-1){400}}
\put(-100,+100){\line(+1,-1){400}}
\put(-300,-300){\line(+1, 0){600}}
}
\mbox{\hspace{1cm}}

\vfill

\begin{center}
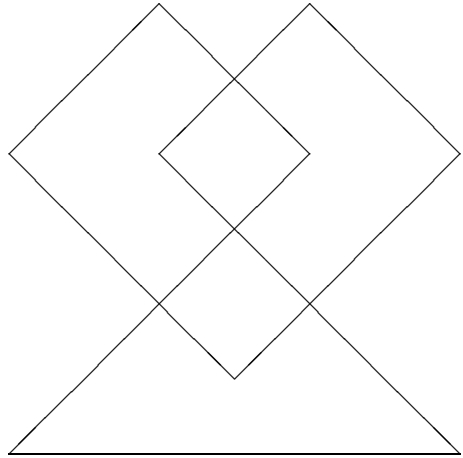
{\bf Fig.~4:}\quad Alternating Whitehead link at $D=-4$
\end{center}
\mbox{\hspace{1cm}}\hfill
\dia{
\put(   0, 400){\line(+1,-2){250}}
\put(   0, 400){\line(-1,-2){250}}
\put(-250,-100){\line(1,0){500}}
\put(-360,000){\line(3,1){720}}
\put(-360,240){\line(3,-1){720}}
\put(-360,240){\line(0,-1){240}}
\put(+360,240){\line(0,-1){240}}
}
\mbox{\hspace{1cm}}

\newpage

\begin{center}
{\bf Fig.~5:}\quad Alternating link $6^3_1$ at $D=-7$
\end{center}
\mbox{\hspace{1cm}}\hfill
\dia{
\put(-400,+400){\line(+1,-2){250}}
\put(+100,+400){\line(-1,-2){250}}
\put(-400,+400){\line(+1, 0){500}}
\put(-100,+300){\line(+1, 0){500}}
\put(-100,+300){\line(+1,-2){250}}
\put(+400,+300){\line(-1,-2){250}}
\put(-350,+50){\line(+1, 0){500}}
\put(-350,+50){\line(+1,-2){250}}
\put(+150,+50){\line(-1,-2){250}}
}
\mbox{\hspace{1cm}}

\vfill

\begin{center}
{\bf Fig.~6:}\quad Alternating link $9^2_{40}$ at $D=-8$
\end{center}
\mbox{\hspace{1cm}}\hfill
\dia{
\put(+400,+400){\line(-1, 0){800}}
\put(+400,+400){\line(-1,-1){600}}
\put(-400,+400){\line(+1,-1){600}}
\put(+200,-200){\line(+1,+1){300}}
\put(-200,-200){\line(-1,+1){300}}
\put(-500,+100){\line( 1, 0){1000}}
\put(-400,+300){\line( 1, 0){800}}
\put(-400,+300){\line( 1,-1){400}}
\put(+400,+300){\line(-1,-1){400}}
}
\mbox{\hspace{1cm}}

\newpage

\begin{center}
{\bf Fig.~7:}\quad Alternating link
$(\sigma_1^{}\sigma_2^{-2}\sigma_3^{}\sigma_2^{-2})^2$ at $D=-11$
\end{center}
\mbox{\hspace{1cm}}\hfill
\dia{
\put(-400,+400){\line(1,-0){800}}
\put(-400,+400){\line(+1,-1){400}}
\put(+400,+400){\line(-1,-1){400}}
\put(-400,-100){\line(1,-0){800}}
\put(0,300){\line(+1,-1){400}}
\put(0,300){\line(-1,-1){400}}
\put(-100, 0){\line(0,+1){300}}
\put(+100, 0){\line(0,+1){300}}
\put(-100, 0){\line(+1,+1){150}}
\put(+100, 0){\line(-1,+1){150}}
\put(-100,300){\line(+1,-1){150}}
\put(+100,300){\line(-1,-1){150}}
}
\mbox{\hspace{1cm}}

\vfill

\begin{center}
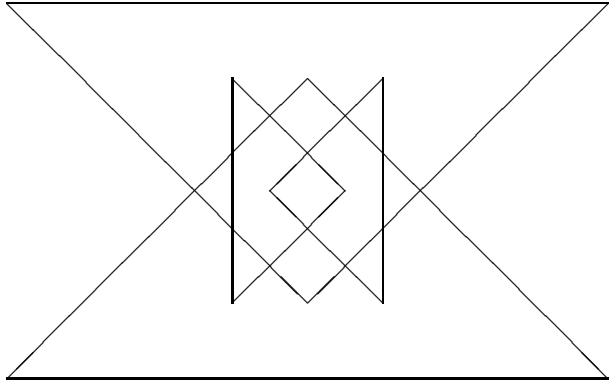
{\bf Fig.~8:}\quad Alternating link
$(\sigma_1^2\sigma_2^{-2})^3$ at $D=-15$
\end{center}
\mbox{\hspace{1cm}}\hfill
\dia{
\put(   0, 400){\line(+2,-3){400}}
\put(   0, 400){\line(-2,-3){400}}
\put(-400,-200){\line(1,0){800}}
\put(   0,-400){\line(-2,+3){400}}
\put(   0,-400){\line(+2,+3){400}}
\put(-400, 200){\line(1,0){800}}
\put(-300,-150){\line(1,0){600}}
\put(-300,-150){\line(+2,+3){300}}
\put(+300,-150){\line(-2,+3){300}}
}
\mbox{\hspace{1cm}}

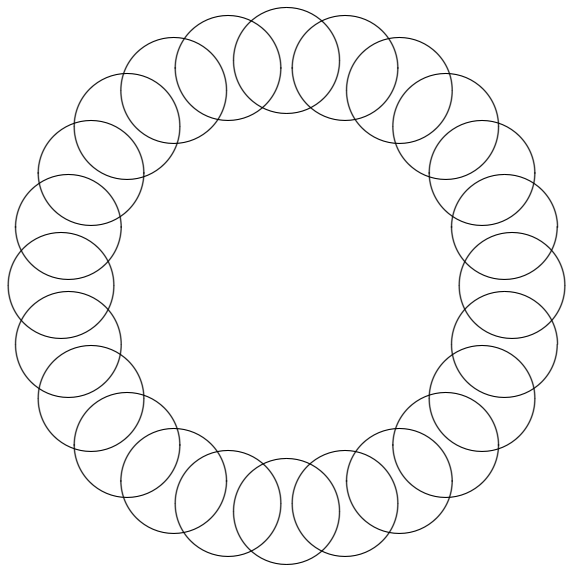
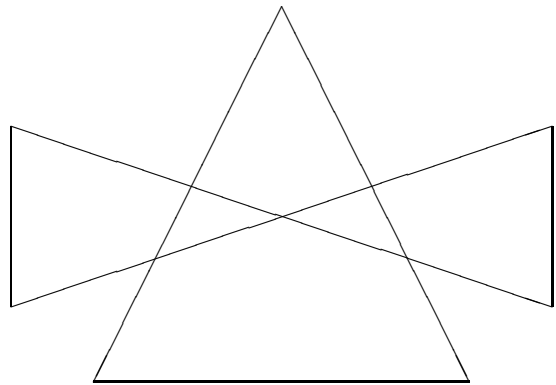
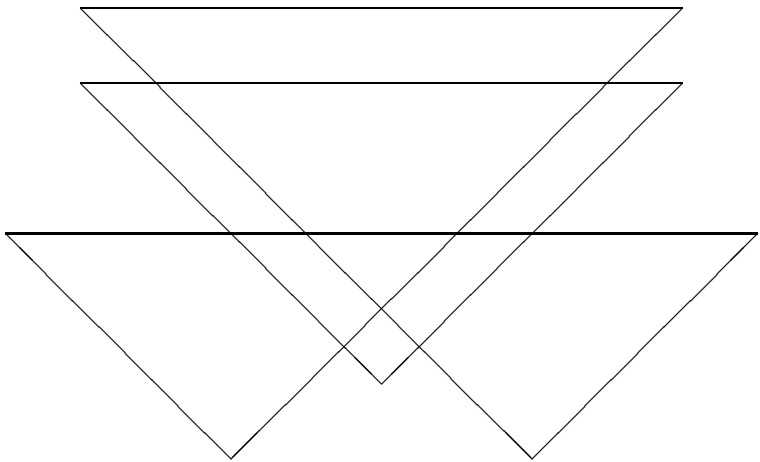
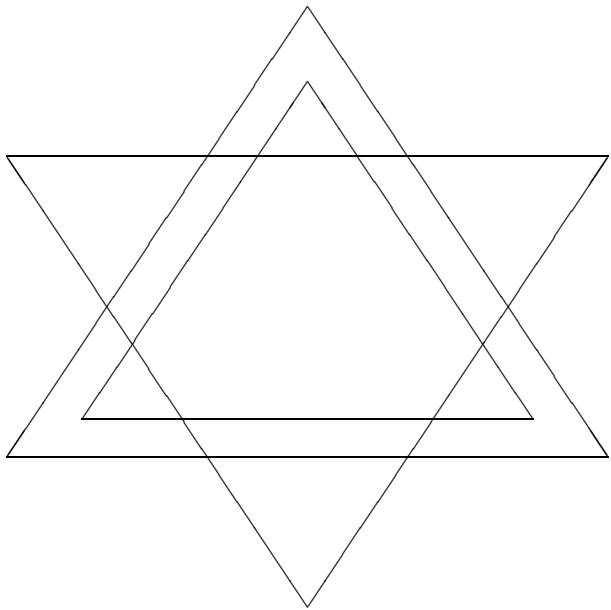
\end{document}